\documentclass[twocolumn, superscriptaddress]{revtex4}

\usepackage{xcolor}
\usepackage{graphicx}  % needed for figures
\usepackage{epstopdf}
\usepackage{dcolumn}   % needed for some tables
\usepackage{bm}        % for math
\usepackage{amsmath}
\usepackage{amssymb} 
\usepackage{mathtools}
\usepackage[normalem]{ulem}
\usepackage{enumerate}

\newcommand{\K}{\Box}
\newcommand{\W}{\Diamond}

\begin{document}

\author{Michael E. Beverland}
\affiliation{Station Q Quantum Architectures and Computation Group, Microsoft Research, Redmond WA, USA}
\author{Benjamin J. Brown}
\affiliation{Niels Bohr International Academy, Niels Bohr Institute,  Copenhagen, Denmark}
\affiliation{Centre for Engineered Quantum Systems, School of Physics, University of Sydney, Sydney, New South Wales 2006, Australia}
\author{Michael J. Kastoryano}
\affiliation{Niels Bohr International Academy, Niels Bohr Institute,  Copenhagen, Denmark}
\affiliation{Institute for Theoretical Physics, University of Cologne, Germany}
\author{Quentin Marolleau}
\affiliation{D\'{e}partement de Physique, \'{E}cole Normale Sup\'{e}rieure, Paris-Saclay}

\title{The role of entropy in topological quantum error correction}

\begin{abstract}
The performance of a quantum error-correction process is determined by the likelihood that a random configuration of errors introduced to the system will lead to the corruption of encoded logical information. 
In this work we compare two different variants of the surface code with a comparable number of qubits: the surface code defined on a square lattice and the same model on a lattice that is rotated by $\pi / 4$.
This seemingly innocuous change increases the distance of the code by a factor of $\sqrt{2}$. 
However, as we show, this gain can come at the expense of significantly increasing the number of different failure mechanisms that are likely to occur. 
We use a number of different methods to explore this tradeoff over a large range of parameter space under an independent and identically distributed noise model. 
%\mb{Using an analytical model that generalises path counting methods we find strong evidence that the rotated model will outperform the square lattice model at asymptotically large system sizes, and at fixed sub-threshold error rates.
%Remarkably, using intensive numerics, we also find a region of parameter space below the threshold error rate where the square lattice surface code marginally outperforms the rotated model. We find this regime persists for systems up to sizes of the order of around two thousand qubits.}{{\bf I actually don't think the previous two sentences emphasize all our key findings. How about:} 
We rigorously analyze the leading order performance for low error rates, where the larger distance code performs best for all system sizes.
Using an analytical model and Monte Carlo sampling, we find that this improvement persists for fixed sub-threshold error rates for large system size, but that the improvement vanishes close to threshold.
Remarkably, intensive numerics uncover a region of system sizes and sub-threshold error rates where the square lattice surface code marginally outperforms the rotated model. 
\end{abstract}
\maketitle

 \section{Introduction}
 
The physics of the quantum error correction process~\cite{Shor95, Steane96, Kitaev03, Dennis02, Terhal15, Brown16}, where we aim to correct a random configuration of errors incident to a system, is naturally captured by its free energy; a quantity specified by an energetic and an entropic contribution. Broadly speaking, the likelihood that the environment can introduce an error that will corrupt encoded logical information corresponds to the configurational energy, and the number of configurations which will lead the system to failure -- the entropic contribution to the free energy. Characterising both of these quantities will enable us to establish the performance of a given quantum error-correction procedure precisely.

A number of studies have been conducted where threshold error rates of topological codes have been determined under a variety of noise models by mapping maximum likelihood decoding onto the partition function of a statistical mechanical Hamiltonian. Under such mappings thresholds have been found to correspond to phase transitions in the statistical mechanical model which can be estimated both analytically~\cite{Dennis02, Fowler12} and numerically~\cite{Katzgraber10, Andrist11, Bombin12, Andrist12, Katzgraber13, Andrist15, Andrist16, Kubica17, Chubb18}. More generally, it is interesting to explore how the competition between energy and entropy is manifest at low error rates. For instance, such studies~\cite{Pastawski10} have shown that certain models with a divergent code distance such as the Bacon-Shor code~\cite{Bacon06} do not present a finite error threshold due to entropic considerations that arise over the error-correction procedure. Indeed, we hope that we can develop quantum hardware~\cite{Reed12, Barends14, Nigg14, Corcoles15, Takita16, Linke16} that functions with error rates far below threshold such that the logical failure rate of the system decays rapidly as we increase the number of physical degrees of freedom. To this end it is important to characterise the energetic and entropic contribution of the free energy of different quantum error correction processes to better understand the shortcomings of their performance, and to help us design better quantum error-correcting codes in the future.

\begin{figure}
\includegraphics{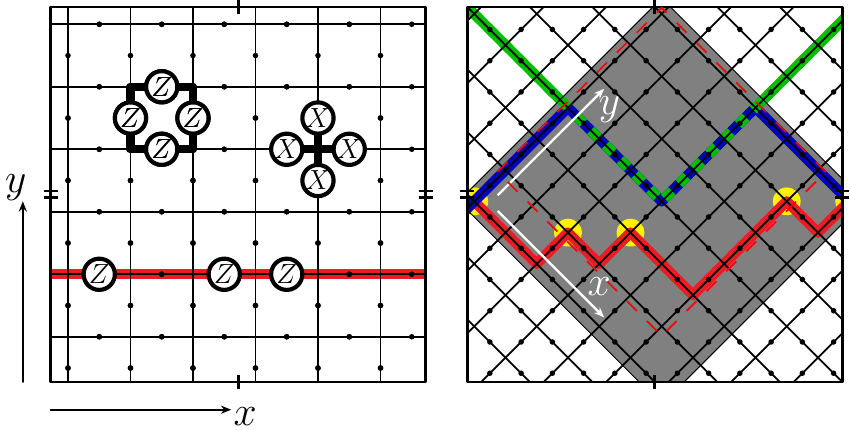}
\caption{\label{Fig:Lattices}Square lattice surface codes on the torus with two different orientations. (Left) The square-lattice surface code with $n = 72$ and $d = 6$. A star and plaquette operator are shown at the top of the figure. A least-weight error is shown along the red path at the bottom of the lattice. The red line supports an uncorrectable error of dephasing flips with weight $d/2$. (Right) The rotated diamond-lattice surface code with $n = 144$ and $d = 12$. The coloured lines indicate the low-weight support of different low-weight logical operators, and the yellow points indicate right turns in the red path.}
\end{figure}

We present new insights into the statistical mechanics of topological quantum error correction by exploring the following observation, depicted in Fig.~\ref{Fig:Lattices}. The surface code~\cite{Kitaev03, Dennis02}, first defined by Kitaev with qubits lying on the edges of a square lattice has a code distance of $d = \sqrt{n / 2}$ where $n$ is the number of physical qubits in the system. In contrast, ~\cite{Bombin06a, Hastings15, Delfosse16} the model~\cite{Wen03} where the same lattice geometry is rotated by $\pi / 4$ ~\cite{Nussinov09, Brown11} has an increased code distance; $d = \sqrt{n}$. Conventional wisdom may lead us to suppose that that the rotated variant of the surface code will have a better qubit economy than the original square lattice model. However, to compare these two codes accurately, we must account for intricate combinatorial effects that emerge over the error-correction process. In fact, we find that the better logical error rate of these two models is a complicated function of their system size and error rate.

\begin{figure}
	\includegraphics{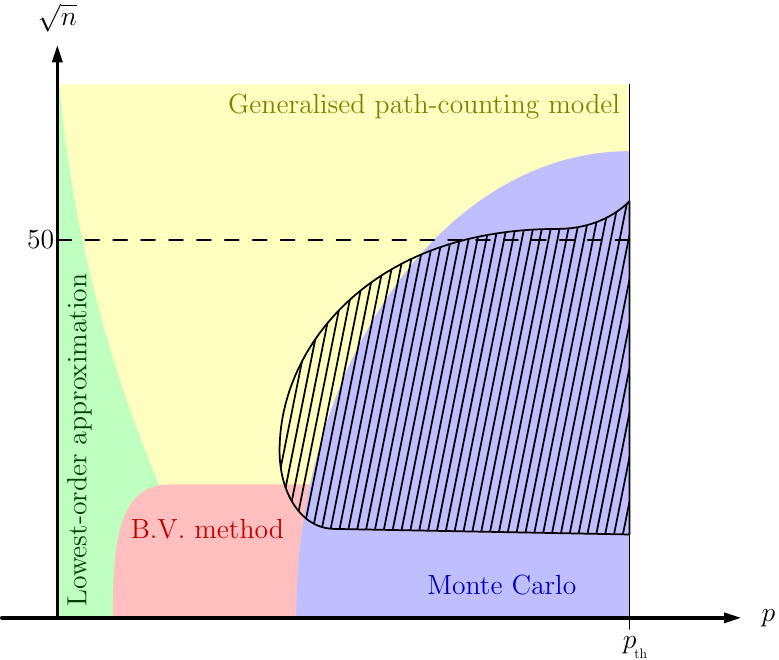}
	\caption{Summary of results. The failure probability $\overline{P}(p,n)$ of the two orientations depends on the number of qubits $n$, and physical error rate $p$. Surprisingly, we find at error rates greater than about half the threshold error rate $p_\text{th}$ for system sizes $\sqrt{n} \lesssim 50 $, that the code with smaller distance outperforms the rotated variant due to entropic effects.
		We hatch this region in parameter space. We use a number of approaches to explore the different parts of parameter space that are shown with different colors in the figure.
		In Sec.~\ref{sec:pathcounting} we find closed expressions for $p \rightarrow 0$ (green). 
		For large $n$, these have the form $\overline{P}(p,n) \rightarrow (\gamma_0)^{\sqrt{n}} p^{d/2} $, with $\gamma^{\K}_0 = 2^{\frac{1}{\sqrt{2}}} \approx 1.6325$ and $3.4142 \approx 2+ \sqrt{2} \leq \gamma^{\W}_0 \leq \sqrt{27/2} \approx 3.6742$. In Sec.~\ref{sec:MC} we use Monte Carlo sampling to study the finite error rate regime numerically (blue). We use the splitting method (red), introduced for topological codes by Bravyi and Vargo in Ref. ~\cite{Bravyi13}, to numerically interpolate between the Monte-Carlo studies and the analytic path counting results which verifies our analysis in the low error rate regime. Finally, in Section~\ref{sec:model} we introduce an analytical model that can be regarded as a generalisation of the path-counting method that accurately models a number of features seen in the numerical studies (yellow). 
		\label{fig:summaryintro}
}
\end{figure}

In Fig.~\ref{fig:summaryintro}, we summarize the results of the paper on a diagram that maps out parameter space over system size and noise strength, indicating which analysis tools apply to what regimes, and the area of parameter space where we might prefer to employ the code with the inferior distance. 
In Sec.~\ref{Sec:Preliminary} we introduce the surface code, the error model, the decoding algorithm we use for numerical studies, and we define the free energy of the quantum error correction procedure. 
In Sec.~\ref{sec:pathcounting} we study both models analytically in the limit of low error rate, where energy dominates entropy and the rotated code outperforms the original. 
We review the failure rate calculation of the original model~\cite{Dennis02} in this limit and derive a new formula for the rotated model. 
In Sec.~\ref{sec:MC} we use numerical methods to observe that the two codes behave almost identically close to the threshold error rate. Remarkably, for finite system sizes up to $n \lesssim 2500$, we find that the unrotated variant of the code has a marginally lower logical error rate.
We also use sophisticated numerical methods~\cite{Bennett76, Bravyi13} to verify our analysis in the low error rate regime.
In Sec.~\ref{sec:model} we generalise the path counting calculations for the low error rate regime to gain insight into the numerical results for finite error rate.

\section{Error correction with the surface code}
\label{Sec:Preliminary}

We first describe the surface code, the error model and the minimum-weight matching decoding algorithm we use to obtain our results. A detailed description of these concepts are found in the seminal references~\cite{Kitaev03, Dennis02} and in recent review articles~\cite{Terhal15, Brown16}.

\subsection{The surface code}
The surface code is defined on a graph embedded on a topologically non-trivial two-dimensional manifold with qubits on its edges. Encoded states of the surface code, $ | \psi \rangle $, lie in the common $+1$ eigenspace of elements of its stabilizer group, $\mathcal{S} \subseteq \mathcal{P}_n$, where $\mathcal{P}_n$ is the Pauli group acting on $n$ qubits which is generated by the Pauli operators $X_e$, $Y_e$ and $Z_e$ acting on the qubit of edge $e$~\cite{Gottesman97}. The stabilizers of the surface code, commonly known as star and plaquette operators, $S_v, S_f \in \mathcal{S}$, are associated to vertices, $v$, and faces, $f$, of the graph. Specifically, we have star operators $S_v = \prod_{\partial {e}\ni v} X_e$ and plaquette operators $S_f = \prod_{e \in \partial f} Z_e$, where $\partial e$($\partial f$) denote the set of vertices(edges) in the boundary of $e$($f$). See Fig.~\ref{Fig:Lattices} for examples of star and plaquette operators.

The logical operators that generate the rotations about the encoded states of the surface code, $\overline{Z}_j$ ($\overline{X}_j$), are the tensor product of Pauli-Z(Pauli-X) operators supported along non-trivial cycles of the graph(dual-graph). Examples of non-trivial cycles of the graph are shown by colored lines in Fig.~\ref{Fig:Lattices}. Closed cycles of Pauli-operators that form contractable loops are elements of the stabilizer group.

Important to our study is the code distance, $d$ which is the weight of the least-weight non-trivial logical operator. In this paper, we only consider codes with even distance. The lattice shown on the left of Fig.~\ref{Fig:Lattices}, i.e. the square-lattice code, has a code distance $d = \sqrt{n/2}$, and the lattice on the right, the rotated-lattice model, has code distance $d = \sqrt{n}$, where $n$, the code length, is the number of qubits in the system. Where there is ambiguity, we use symbols `$\K$' and `$\W$' to distinguish the square(rotated) lattice surface codes shown to the left(right) of Fig.~\ref{Fig:Lattices}. To the best of our knowledge, the first appearance of the surface code with the square(rotated)-lattice was due to Kitaev(Wen) in Refs.~\cite{Kitaev03} and~\cite{Wen03}, respectively.

\subsection{Error model}
\label{sec:errormodel}
Stabilizer measurements project local noise onto an encoded state that has been acted upon by some Pauli error. We thus focus our attention on Pauli errors $E \in \mathcal{P}_n$.
Given that the surface code identifies Pauli-Z(Pauli-X) errors separately in an equivalent manner, and the group $\mathcal{P}_n$ is generated by operators $X_e$ and $Z_e$, we can exclusively study Pauli-Z errors that are identified by star operators. An equivalent discussion holds for Pauli-X errors that are identified by plaquette operators. We therefore consider errors of the form 
\begin{equation}
E = \prod_{e\in\mathcal{Q}} Z_e^{x_e}\label{eqn:E},
\end{equation}
where $x_e = 1$ with constant probability $p$ and $x_e = 0$ otherwise. We can concisely write the probability that an error, $E$, is drawn from the independent and identically distributed dephasing noise model $\pi(E)$ with the expression
\begin{equation}
\pi (E) = (1-p)^{n - \text{wt}(E)} p ^{\text{wt}(E)}=(1-p)^{n}e^{-\beta \text{wt}(E)}, \label{Eqn:iidDistribution}
\end{equation}
where $\text{wt}(E) = \sum_e x_e$ is the weight of error operator $E$. It will also be helpful to denote by $\pi_w$ the probability that an error of weight $w$ occurs, i.e. $\pi_w = \pi(E)$ for $\text{wt}(E) = w$. 

%{\color{blue}  We additionally denote the Hamming weight of a vector with notation $|\vec{x}|$ which denotes the number of elements of $\mathbf{x}$ that take the value 1. With this }

\subsection{The minimum-weight matching decoder}

\label{SubSec:Decoder}

Let $| \phi \rangle = E |\psi \rangle$ be a code state which has suffered error $E$. %The error is identified by measuring the stabilizer operators of the code. 
For the surface code, it is helpful to interpret errors as strings that lie along edges of the lattice.
We measure defects, i.e. stabilizer measurements that return the $-1$ outcome, to  determine a correction operator $C$ where $ CE \in \mathcal{S}$ such that $C | \phi \rangle = |\psi \rangle$. The defects from the stabilizer measurements are associated with end points of error strings. Assuming $E$ is low weight, we seek to find an operator $C$ that connects pairs of defects via strings of low weight. We evaluate $C$ with minimum-weight matching~\cite{Edmonds65, Kolmogorov09, Dennis02}. The algorithm takes a complete graph where vertices represent defects and the edges are assigned weights according to the separation of the defects by Manhattan distance. The algorithm returns a bipartite graph whose edges correspond to a least-weight correction.

It is instructive to view error correction with the surface code from a homological perspective. Given that $C$ and $E$ must have the same boundary that is indicated by the locations of stabilizer defects, the product of $CE$ necessarily forms a closed cycle on the graph. Error correction fails if and only if $CE$ produces a non-trivial cycle as this operator will correspond to a logical operator. Otherwise we have that $CE \in \mathcal{S}$ such that error correction does not introduce a logical error to the system. For an introduction to homology theory see Ref.~\cite{Nakahara} or Appendix A of Ref.~\cite{Anwar14} for an introduction that connects topological error correction with homology.

\subsection{The entropic contribution to failure rates}
\label{sec:entropy}

We analyse and compare the the logical failure rates of the square-lattice and rotated-lattice surface codes. We denote the logical failure probability
\begin{equation}
\overline{P}(p,n) = \sum_{E \in \mathcal{F}} \pi(E), 
\end{equation}
where $\mathcal{F}$ is the set of all errors that will cause the logical failure of a code and we append a superscript $\K$ or $\W$ symbol where there is ambiguity which model we are discussing.

By expressing the likelihood of error $\pi(E)$ with $\text{wt}(E) = w$ occurring as $ \pi(E) = (1-p)^n \exp(-\beta w)$ where $\beta = \log\left[(1-p) / p \right]$ plays the role of an inverse temperature, we can express the logical failure rate as follows
\begin{equation}
\overline{P}(p,n) = (1-p)^{n} \sum_{w=d/2}^{n} N_{\text{fail}}(w) e^{ -\beta w},
\end{equation}
where the $N_{\text{fail}}(w)$ is the number of weight-$w$ elements of $\mathcal{F}$. We remark that the summation begins at $w = d / 2$ as in the model of interest this is the least-weight error that will cause the minimum-weight matching decoder to fail. Finally, by defining entropy $S_{\text{fail}} = \log[N_\text{fail}(w)]$ we have the logical failure rate in terms of free energy such that
\begin{equation}
\overline{P}(p,n) = (1-p)^{n} \sum_{w=d/2}^{n} e^{-\beta F(w)},
\label{eq:entropy}
\end{equation}
where
\begin{equation}
F(w) =  w - S_\text{fail}(w) / \beta,
\end{equation}
and the weight of the error $w$ can be regarded as the energy term of the free energy.

Ubiquitous in the behaviour of physical systems is the competition between the energy term and the entropy term of their free energy. 
For a given $\beta$, there will be an energy $\langle w(\beta) \rangle$ which minimizes the free energy $F$, and this dominates the performance $\overline{P} \sim (1-p)^n e^{-\beta F(\langle w(\beta) \rangle)}$.
For two different error correction schemes, the scheme which has the largest minimum free energy will have better performance.
As we have already discussed, the distance of the surface code embedded on the rotated lattice is greater than that of the square lattice model, and therefore has an energetic advantage. 
However, the difference in the performance of the two models becomes less clear when the entropy term is also taken into account. 
As we summarise below, we use a number of different methods to compare the entropy for the square-lattice and diamond lattice surface-code models as a function of the available number of physical qubits and the physical error rate.

%We begin with a path-counting analysis. Here we focus on the limit where $p \rightarrow 0$ such that the relevant quantity to determine the logical failure rate is simply $N_\text{fail}(d/2)$. As we might expect, the logical failure rate of the code with the larger distance has the superior performance in this limit. Curiously, the path-counting analysis also indicates a regime below threshold where the square lattice model outperforms its counterpart. To explore this crossing point we turn to other methods.

%{\color{blue} I think we need to elaborate the following sentences. There is also a diagram we all have in mind, `the phase diagram', to help this description.}
%For fixed $n$ and asymptotically small $p$ (large $\beta$) the smallest energy $w=d/2$ dominates, which we analyze analytically in Section~\ref{sec:pathcounting}.
%We model the case of finite $p$ analytically in Section~\ref{sec:model}.We study all both regimes numerically in Section~\ref{sec:MC}.

\section{Low physical error rates}
\label{sec:pathcounting}

We first consider the regime of asymptotically low error rate. In this regime we can restrict our consideration to logical failures caused by errors of minimal weight, $d/2$. Specifically, we take the following limit
\begin{equation}
  \overline{P}_{\text{low-}p}(p,n) : =  \lim_{p \rightarrow 0} \overline{P}(p,n) \sim N_\text{fail}(d/2) ~p^{d/2}.
\end{equation}
%where we have omitted the prefactor of $(1-p)^n$ on the right-hand side of the equation. 
We now focus on finding analytical expressions for $N_\text{fail}(d/2)$ for each of the models. %The regime where the error rate is sufficiently low that this approximation is valid is often referred to as the `path-counting regime' due to the nature of the calculation we now elaborate upon in this Section.

\subsection{The square lattice model at low error rates}

Path counting has already been well-studied with the square-lattice variant of the surface code~\cite{Dennis02, Watson14} where the code distance is $d= \sqrt{n / 2}$. Logical failures can occur when the error incident to the system is supported on at least half of the qubits of a horizontal or a vertical path through the lattice. There are $2d$ such paths; we illustrate one of them in red to the left of Fig.~\ref{Fig:Lattices}. 
We can therefore count the number of least-weight errors by multiplying the number of minimal-weight non-trivial cycles by the number of error configurations of weight $d/2$ that lie on each given cycle. The number of configurations on a given path is just the binomial coefficient $ C^{d}_{d/2}$ where $C^a_b = a! /b!(a-b)!$, yielding
\begin{eqnarray}
N_{\text{fail}}^{\K}(d/2) &= & \frac{1}{2} \cdot  2d \cdot C^{d}_{d/2}.
\label{eq:NpcK}
\end{eqnarray}
The factor of $1/2$ comes from the fact that we have assumed $d$ is an even integer whereby each syndrome will occur for exactly two possible errors, $E_1$ and $E_2$, such that $E_1E_2$ is a non-trivial logical operator. Supposing then that the decoder is deterministic, and the correction operator for the given syndrome is,  say $C = E_1$, then  the decoder will decode this syndrome correctly in half of the instances where this syndrome appears under the error model. Indeed, on average, half of the times this syndrome occurs will be due to incident error $E_1$, and the other half will be due to error $E_2$.

In the limit of large $d$ we can make use of Stirling's approximation to obtain $C^d_{d/2}\approx \sqrt{\frac{2}{\pi d}}2^d$. 
Dropping terms polynomial in $n$ yields the large $n$ expression
\begin{eqnarray}
N_{\text{fail}}^{\K}(d/2) &\sim & (\gamma^{\K}_0)^{\sqrt{n}},
\end{eqnarray}
where
\begin{eqnarray}
\gamma^{\K}_0 = 2^{\frac{1}{\sqrt{2}}} \approx 1.6325.
\end{eqnarray}

\subsection{The rotated lattice at low error rates}
\label{subsec:rotatedpathcounting}

It is instructive to compare the rotated code to the square lattice model in the low error rate regime because we observe that, although the rotated lattice has an increased distance, $d^{\W} =  \sqrt{2} d ^\K$, it also has a considerably larger number of least weight errors. Indeed, consider moving from the left point to the right point of the gray diamond on the lattice on the right of Fig.~\ref{Fig:Lattices}. For a given starting point in Fig.~\ref{Fig:Lattices}, there are  $C^{d}_{d/2}$ distinct paths within the diamond. 
Given that up to  $C^{d}_{d/2}$ errors can be arranged  along each path, we find an upper bound on the number of least-weight errors $N^\W_\text{fail}(d/2)$ on the rotated lattice to be $\sqrt{n} \cdot 2^{2 \sqrt{n}}$.
Before launching into an involved calculation of $N^\W_\text{fail}(d/2)$ we consider some general arguments.

We remark that there are also logical error paths of length $d$ that wrap around a diagonal of the lattice. 
We focus only on the paths that wrap horizontally or vertically around the lattice as there are significantly more of these, and for large $n$ the contribution of the failures due to $d/2$ errors lying on these diagonal paths is negligible by comparison.

This upper-bound over-estimates the number of least weight errors that will cause a logical failure with the rotated lattice. To see why this is the case, recall that we are interested in the number of least-weight errors, not the number of least-weight paths that can support a least-weight error.

Some of the paths we have considered in the calculation so far can overlap, see for instance, the blue and green paths at the right of Fig.~\ref{Fig:Lattices}. 
Indeed, with the counting given above, we have counted an error of weight $d/2$ that lies entirely on both the blue path and the green path twice. In fact, this same error could have appeared on any of $C^{d/2}_{d/4}$ different paths. Supposing that all the errors have been over counted $ C^{d/2}_{d/4} \sim 2^{\sqrt{n} / 2}$ times, we can lower bound the number of least weight errors to $n^{1/4} \cdot 2^{3 \sqrt{n} / 2}$, where we have divided by the largest possible number of non-trivial cycles that every error may have come from by which we may have over counted. 
To this end, we anticipate a large $n$ expression for the number of failing configurations
\begin{eqnarray}\label{eqn:pc2}
N_{\text{fail}}^{\W}(d/2) &\sim & (\gamma^{\W}_0)^{\sqrt{n}},
\end{eqnarray}
where $2.8284 \approx 2 \sqrt{2} \le \gamma^{\W}_0 \le 4$.
In Sec.~\ref{sec:pathcountingWen} and in the associated Appendix~\ref{app:pc} we tighten these bounds on $N_{\text{fail}}^{\W}(d/2)$ giving
\begin{eqnarray}
3.4142 \approx 2 + \sqrt{2} \leq \gamma^{\W}_0 \leq \sqrt{27/2} \approx 3.6742.\label{eqn:tighter}
\end{eqnarray} 
These bounds hold for any choice of minimum weight matching decoder for the rotated lattice, although as we describe in Appendix~\ref{app:decoder}, the path counting performance of different decoders can vary. 
We conjecture that the upper bound is tight for any minimum weight decoder.
%We conjecture that this holds for a `typical' choice of minimum weight matching decoder for the rotated lattice.

%\begin{equation}\label{eqn:pc2}
%\overline{P}^{\W}_{\text{p.c.}} \sim \exp\left( \frac{1}{4}\log n + \gamma \sqrt{n} \cdot \log 2 - \frac{1}{2} \beta \sqrt{n} \right),
%\end{equation}
%where $ 3/2 \le \gamma \le 2$ and again, we have neglected a factor of $(1-p)^n$. 

With these ranges of $\gamma^{\K}_0$ and $\gamma^{\W}_0$ we compare the two models by studying the ratio
$  \overline{P}^{\W}_{\text{low-p}} / \overline{P}^{\K}_{\text{low-p}} \sim \Delta(p)^{\sqrt{n}} $ where 
\begin{eqnarray}
\Delta(p) &=& \frac{\gamma^{\W}_0}{\gamma^{\K}_0} \cdot  p^{\frac{d^{\W}-d^{\K}}{2\sqrt{n}}} .\nonumber 
\end{eqnarray}
Written in this form it is clear that in the regime of low $p$,  $\Delta(p)<1$ meaning that the smaller entropic contribution of the square lattice compared with the rotated lattice is not enough to make up for the its smaller distance. 
We therefore observe that the toric code on the rotated lattice has a smaller asymptotic logical failure rate than on the square lattice for the same $n$. 
However, this formula for $\Delta(p)$ is greater than one for $p > 0.7 \%$, suggesting that the square lattice could outperform the rotated lattice in spite of its reduced distance for some values of $p$ below threshold. 
Caution must be taken with this conclusion since we are using path counting results (valid only for asymptotically small $p$) to reason about a finite $p$ regime. 
In Sections~\ref{sec:MC} and \ref{sec:model} we analyze the finite $p$ regime numerically and analytically, and find that this crossover survives for finite size lattices up to $n\sim 2500$.

\subsection{Tighter bounds for path counting with the rotated lattice}
\label{sec:pathcountingWen}

We now set out to derive the tighter bounds of Eqn. (\ref{eqn:tighter}) for the number of minimum weight failing errors $N_{\text{fail}}^{\W}(d/2)$. 
To do so, we define a map from each weight-$d/2$ error configuration either to a weight-$d$ logical operator containing the error, or to `null' when no such logical operator exists.
Then we step through each weight-$d$ logical operator, which can be enumerated straightforwardly, and count the number of errors in its pre-image.
This allows us to enumerate every weight-$d/2$ error which can cause a minimum weight decoder to fail, without counting the same error multiple times (as each error is only in the pre-image of one logical operator).

We start by enumerating all horizontal logical operators, that we will also be refer to as paths, which pass through the left and right corners of the gray diamond depicted to the right of Fig.~\ref{Fig:Lattices}. Note that any minimal weight path passing through these two points will be in the gray diamond. We denote the vertices within the gray area with a coordinate system $(x,y)$ where the origin lies to the left of the figure and the axes are shown in white. 

The principle insight we use to calculate the number of least-weight errors that can lead to logical failure is the characterisation of the paths  by the vertices a given path turns on. Indeed, turns come in two types --- left turns and right turns --- where a left(right) turn is a vertex where the direction of the path changes to move along the $y-$axis($x$-axis). The right turns made by the red path in Fig.~\ref{Fig:Lattices}(right) are marked with yellow spots. Left turns are unmarked.

Given this labeling, we now define a map from a weight-$d/2$ error to a horizontal weight-$d$ logical operators $CE$ containing $E$. 
We define the map by imagining a deterministic decoder that returns a correction $C$ which contains no right turns.
With this particular decoder, we are able to count the number of errors that give rise to a particular path \textit{uniquely} with the following observation: The path $CE$ contains a right turn if and only if there is at least one error lying on the path on an edge adjacent to every vertex that supports that right turn. 
We can therefore count the number of least-weight errors that maps to a particular path by counting the number of configurations where all right turns of a given path are adjacent to an edge that supports an error. It is easy to check that non-trivial cycles of length $d$ are such that every right turn is preceded by a left turn and vice versa.

%We are now in a position to count the number of errors $E$ such that $CE$ gives rise to a particular path.

We count the number of configurations that lie on a path with $T$ right turns by first separating the segments of the path into two sets; those that are adjacent to right turns and those that are not. We first focus on those adjacent to the right turns. Each of the $T$ right turns must have an error on one of its adjacent edges, and we suppose that $0 \le w \le w_\text{max} \equiv \min(T,d/4)$ of the right turns support two errors. Of the set of all right turns the subset of $w$ right turns that are occupied by two errors can be configured in any of $C^T_w$ ways. Now, given that right turns with only one adjacent error can take two configurations determined by which edge the error lies on, and the remaining $w$ right turns with both edges occupied by an error can take only one configuration we find that the edges adjacent to right turns can take $C^T_w \times 2^{T-w}$ configurations if $w$ of the right turns have errors on both of their adjacent edges.

The remaining $d/2 - w - T$ errors that lie on the path can be distributed arbitrarily along the remaining $d - 2T$ edges of the path. We therefore obtain an expression, $\#(T,d)$, for the number of error configurations that give rise to a path with $T$ right turns using this specialised decoder. We find
\begin{equation}
\#(T,d) = \sum_{w = 0 }^{w_{\text{max}}} 2^{T - w} C_w^T C_{d/2 - T - w}^{d - 2T}. \label{Eqn:Turns}
\end{equation}

It remains to count the number of paths around the torus with $T$ right turns. We can specify any path of length $d$ containing $t$ right turns by listing the coordinates of the right turns of the path that lie in the red-dashed square in Fig.~\ref{Fig:Lattices}. We denote these with coordinates $\{(x_1,y_1),(x_2,y_2),\dots,(x_t,y_t)\}$ . 
Every choice of $t$ coordinates such that $x_j$ and $y_j$ take integer values and are strictly increasing (such that $ 0 \le x_1 <  x_2  <  \dots  < x_t \le  d / 2 - 1$ and $ 1 \le  y_1 < y_2 < \dots < y_t \le d / 2$) will specify a valid path.
There are therefore $\left(C_{t}^{d/2}\right)^2$ length-$d$ paths with $t$ right turns from $(0,0)$ to $(d/2,d/2)$.

Finally, to determine the number of paths that include $T$ right turns exactly we must look to the boundary, which may also include a right turn. 
The paths with an extra right turn will be precisely those which do not have $x_1=1$ and do not have $y_T=d/2-1$.
There are $\left( {C_T^{d/2-1} }\right) ^2$ such paths with $T+1$ turns if we include the right turn at the boundary. We therefore subtract this number that accounts for the paths that include an extra right turn at the boundary. To complete this calculation, we then add the terms with $T$ right turns including one on the boundary. These must have $T-1$ turns in the bulk arranged such that $x_1 \not=1$ and $y_{T-1} \not=d/2-1$. We find there are $C^{T-1}_{d-1}$ such paths.
We therefore arrive at $\left( {C_T^{d/2} }\right) ^2 - \left( {C_T^{d/2-1} }\right) ^2 + \left( {C_{T-1}^{d/2-1} }\right) ^2$ different paths with $T$ right corners.
We finally sum over paths with $ 0 \le T \le d/2$ right turns to obtain at the bound
\begin{eqnarray}
\!  N_{\text{fail}}^{\W}\!\left( \! \frac{d}{2} \right) \!\!  &\leq&  \! d  \sum_{T = 0}^{\frac{d}{2}} \! \left[ \! \left( \! {C_T^{\frac{d}{2}} } \! \right)^{\!\!2} \!\! - \! \left( \! {C_T^{\frac{d}{2}-1} }\right)^{\!\!2} \!\! +\! \left( \! {C_{T-1}^{\frac{d}{2}-1} }\right)^{\!\!2} \right] \! \#(T,d)+ \nonumber\\
&&+ ~d \cdot C^d_{d/2} - 2d^2. \label{eqn:pcW}
\end{eqnarray}
The sums pre-factor $d$ accounts for the number of initial points a path can begin from, where we have considered non-trivial cycles along both the horizontal and vertical directions, and the function $\#(T,d)$ is given in Eqn.~(\ref{Eqn:Turns}).
The term $d \cdot C^d_{d/2}$ accounts for the errors which can fail to diagonal paths. 
Note however that there are $d^2$ errors that have been counted three times that run along $d/2$ contiguous edges with a common direction. 
We have therefore subtracted $2d^2$ paths to account for this.

In Appendix~\ref{app:pc} we consider Eqn.~(\ref{eqn:pcW}) in the large $n$ limit and calculate $\gamma^{\W}_0 \leq \sqrt{27/2} \approx 3.6742$ in the expression $N_{\text{fail}}^{\W}(d/2) \sim (\gamma^{\W}_0)^{\sqrt{n}}$.
It is possible to further tighten the upper bound in Eq.~(\ref{eqn:pcW}) by making use of the fact that any minimum-weight decoder must correct at least one error for a given syndrome, but this does not change the bound on $(\gamma^{\W}_0)$.
In the same appendix we also prove the lower bound
$\gamma^{\W}_0 \geq 2 + \sqrt{2} \approx 3.4142$.
We believe the upper bound to be tight.

%\cc{[I think the rest of this subsection needs some work. We need to state (1) that we consider only horizontal logical failures, (2) that we make the decoder unique (given any logical failures are horizontal) by preferring connections along lower paths, (3) that vertical errors can be counted in the same way, even though we have already fixed the decoder to make it unique for horizontal errors (I am not 100\% sure this is true), (4) we need to add the diagonal error contribution, or remove the equality for our sum giving $N_{\text{p.c}}^{\W}$, (5) be more explicit that we use "internal" turns to specify the path, but "total" turns to calculate whether or not a sprinkling of errors on a path will correct to the path.]}

\section{A numerical study}
\label{sec:MC}

Using a low-$p$ approximation we have predicted that for $p \gtrsim 0.4 \%$ it pays to minimise the entropy of the code, and not its distance. However, this prediction is unreliable since it is based on results that only rigorously hold in the limit $p\rightarrow 0$. We next turn to numerical methods to determine the failure rates of the two models at higher error rates, up to threshold. In what follows we describe our simulations. We find that close to threshold, both models behave almost identically for a given $n$. This is in stark contrast with the low-$p$ behavior. Furthermore, we identify a region where the original lattice marginally outperforms the rotated model. %marked by the hatched region in Fig.~\ref{fig:summary}. 
We find that this region does not persist for system sizes greater than $\sqrt{n} \sim 50$ or for error rates below $p \sim 0.05$. We also use the method due to Bravyi and Vargo~\cite{Bravyi13} to verify our low-$p$ formulas given in the previous section. Finally, we fit our data to a two paramter ansatz, which helps characterise the behavior of the two orientations for small system sizes. Our numerical results are summarised in Fig.~\ref{fig:summaryintro}.

Monte-Carlo sampling proceeds by generating $\eta$ instances of error operators $E$ to estimate
\begin{equation}
\overline{P} = 1-\text{prob}(CE \in \mathcal{S}).
\end{equation}
where errors $E$ are drawn from the distribution determined by the error model defined above for a given $p$, and the correction operator $C$ is evaluated using the minimum-weight matching decoder for the syndrome of each error.

\begin{figure}
\includegraphics[width=\columnwidth]{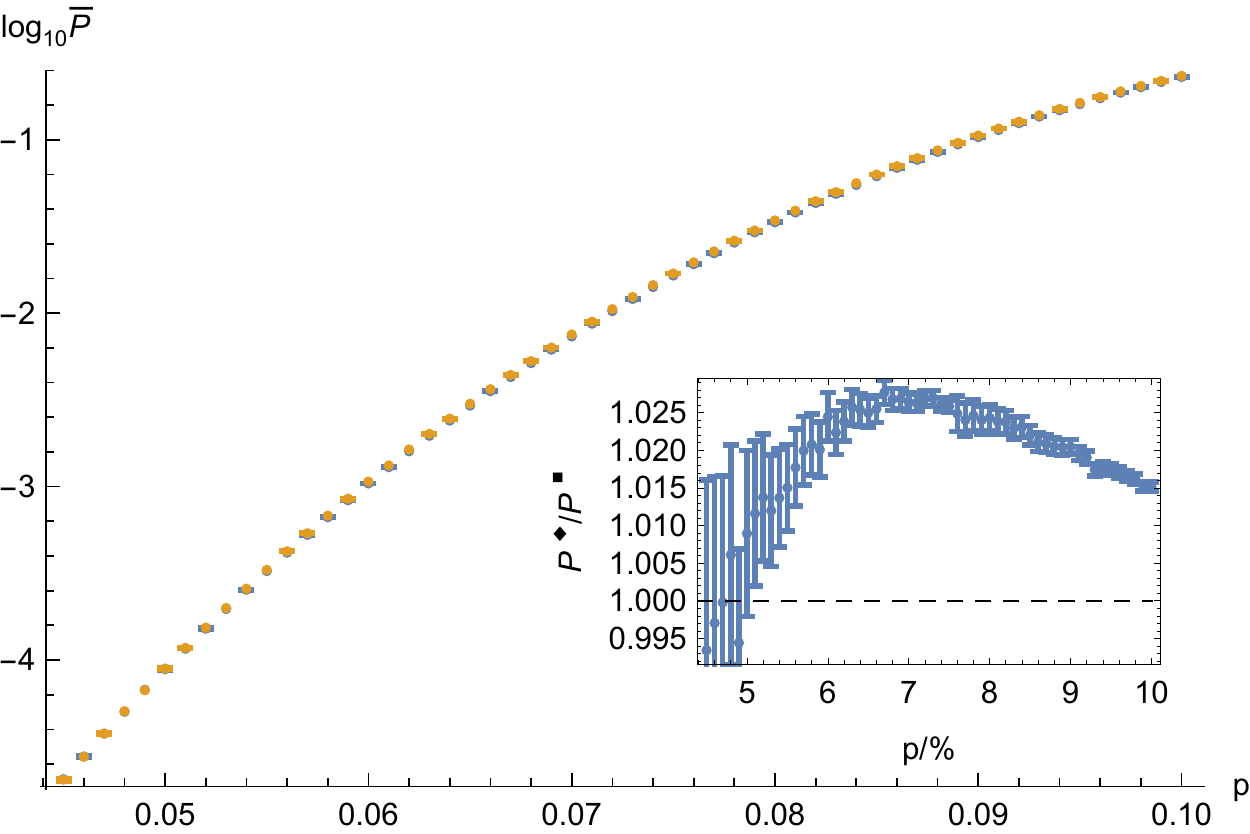}
\caption{\label{Fig:MagicNumber} The logical failure rate for the square lattice model with $n^\K = 1152 (d^\K = 24)$, shown in blue, compared with that of the rotated lattice with $n^\W = 1156 (d^\W = 34)$ in yellow, calculated using $\eta \sim 10^8 \text{--} 10^9$ samples. The inset shows the ratio of the failure rates of the two models, where the ratio in excess of unity marks the region where the square lattice model outperforms the rotated model using four fewer qubits.}
\end{figure}

We begin by identifying a point where the original lattice outperforms the rotated lattice. In Fig.~\ref{Fig:MagicNumber} we compare the logical failure rates of the unrotated model with distance $d^\K = 24$ and the rotated model with $d^\W = 34$ with Monte-Carlo sampling. We choose these two system sizes as they each use a similar number of qubits, $n^\K = 1152$ compared with $n^\W = 1156$. Error bars are determined according to $\Delta \overline{P} = \sqrt{(1-\overline{P}) \overline{P} / \eta}$ where we collect $ \eta \sim 10^8 \text{--} 10^9$ samples. In the inset we take the ratio of the error rates of the two models, $\overline{P}^\W / \overline{P}^\K$. 

Although the logical failure rate is almost identical for both models, remarkably, the inset shows that the ratio exceeds $1$ at around $p \sim 5\%$. Not forgetting that the two models are equivalent up to their orientation, this is surprising result given that the original lattice has a smaller distance and four fewer qubits than the rotated lattice. We have thus identified a location in parameter space where the distance is not the determining feature of the logical failure rates. We mark this area of parameter space with the hatched region in Fig.~\ref{fig:summaryintro}. It now remains to explore the extent of this behaviour. In what follows we look to determine the boundaries of this region in parameter space.

\subsection{Large system sizes}

\begin{figure}
\includegraphics[width=\columnwidth]{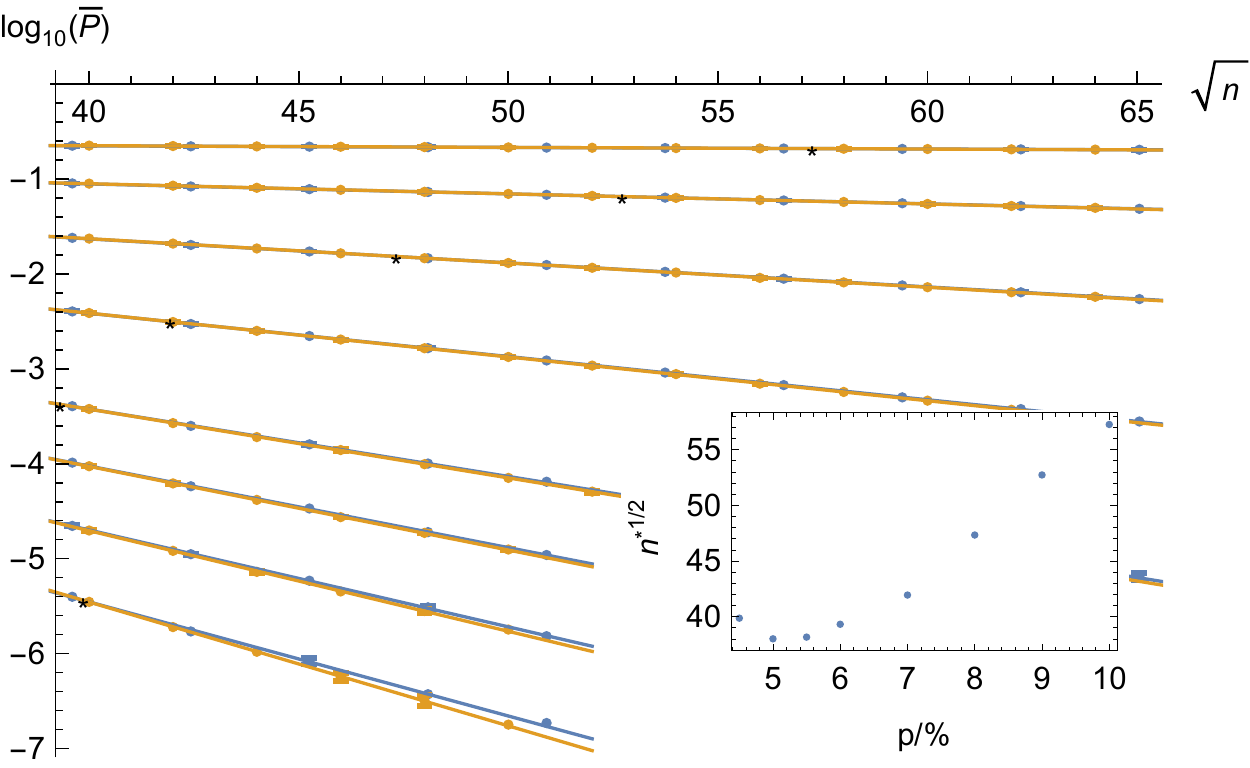}
\caption{\label{Fig:LargeSizes} Monte Carlo data comparing large system sizes of the original(blue) and rotated(yellow) lattice for error rates $p = 4.5\%, 5\%,5.5\%, 6\%, 7\%, 8\%, 9\%, 10\%$ running from the bottom fitting to the top fittings. The inset shows the system size, $L^*$, where the linear fittings of the two different models cross for each value of $p$. These crossing points mark the top of the hatched region in Fig.~\ref{fig:summaryintro}, above which, the rotated lattice begins to outperform the original square lattice model.}
\end{figure}

We next use Monte-Carlo sampling to look at large system sizes at error rates $p \gtrsim 4.5 \% $ to determine the relative performance of the two models as $n$ diverges. We find that at system sizes larger than $\sqrt{n} \sim 50$ the rotated model again outperforms the square-lattice model. In Fig.~\ref{Fig:LargeSizes} we show logical failure rates for system sizes with $ 40 \lesssim \sqrt{n} \lesssim 64$ for physical error rates $p = 4.5\%,\, 5\%,\,5.5\%,\, 6\%,\, 7\%,\, 8\%,\, 9\%,\, 10\%$ where the smallest error rates are shown by the steep straight lines fitting at the bottom of the figure, and the square(rotated) lattice data and fittings are shown in blue(yellow). Like in Fig.~\ref{Fig:MagicNumber}, the data in Fig.~\ref{Fig:LargeSizes} shows that the performance of the two models are almost indistinguishable. The separation in the two models becomes more apparent when $p$ is small. Indeed, the difference in the gradients of the fittings is appreciable at $p = 4.5\% $, but this difference rapidly vanishes as $p$ increases.

To determine the smallest system size above which the rotated model outperforms the square lattice model we find the system size at which the linear fittings shown in the graph cross. We mark the crossing points with small black crosses in the main plot, and the inset shows the crossing points as a function of $p$. The crossing points numerically estimate the location of the top boundary of the hatched region shown in Fig.~\ref{fig:summaryintro}.

The  numerical results that we have described to this point indicate that, in contrast to the regime of low error-rates given in the previous section, the behaviour of the two models is very similar for $p \ge 5\%$ with a region of parameter space where the square lattice model slightly outperforms the rotated model. To understand the extent of the difference between the two models it is illuminating to extrapolate the fitting found in Fig.~\ref{Fig:LargeSizes} to get an optimistic sense of the magnitude of the difference between the two models using our idealised error model. We use the following Ansatz~\cite{Brown15} to fit the data in the figure
\begin{equation}
\overline{P}_{\text{Ansatz}} = A(p) \exp \left( \alpha(p) \log\left( \frac{p} { p_{\text{th}}} \right) \sqrt{n} \right), \label{Eqn:MCFittingAnsatz}
\end{equation}
where $\alpha(p)$ and $A(p)$ are free parameters that depend on $p$ and $p_{\text{th.}}$ is the threshold error rate. Unsurprisingly, we find identical threshold error rates
\begin{equation}
p_{\text{th.}}^\K \sim 0.1035\pm 0.0002, \quad p_{\text{th.}}^\W \sim 0.1035 \pm 0.0002, \label{Eqn:KitaevThresholdEval}
\end{equation} 
for the two models. We evaluate the threshold and we plot $\alpha(p)$ and $\log_{10}A(p)$ for the two models in App.~\ref{App:MonteCarlo}.

As an example, we extrapolate the fittings found above with Eqn.~(\ref{Eqn:MCFittingAnsatz}) at $p = 5\%$. For this error rate we find 
$$
\log_{10}A^\K = -0.61\pm0.04, \quad \alpha^\K= -0.323\pm 0.003,
$$
and
$$
\log_{10}A^\W = -0.47\pm0.04, \quad \alpha^\W= -0.335 \pm 0.003.
$$
With this extrapolation we find that number of qubits at which the logical failure rate of the rotated model is one half of that of the square lattice model at system is $\sqrt{n} \sim 116$. At this point where we have in excess of ten-thousand physical qubits operating at an error rate below half threshold the logical failure rate of the two models is of the order $\sim 10^{-12}$ which is a relevant error rate for large-scale quantum algorithms~\cite{Fowler12b}. On the logarithmic scale that we use to measure failure rate, a factor of one half is relatively inconsequential. Our results thus indicate that, unless we have a very large number of qubits, it may be more valuable to optimise over factors such as the performance of the two different models when performing logical gates~\cite{Brown16a} instead of code distance at high error rates below threshold.

\subsection{Low error rates}

We now verify the calculations made in the previous section for low error rates. We adopt the method of Bravyi and Vargo~\cite{Bravyi13} to probe logical failure rates, $\overline{P}(n, p_0)$, for low physical error rates, $p_0$, that are intractable by regular Monte Carlo sampling. The method proceeds by spliting~\cite{Rubino09} the logical failure rate into a series of ratios $R_j = \overline{P}(n, p_{j}) / \overline{P}(n, p_{j+1}) $ such that
\begin{equation}
\overline{P}(n, p_0) = \overline{P}(n, p_\Lambda)  \prod_{j=0}^{\Lambda-1} R_j,
\end{equation}
where $\overline{P}(n, p_\Lambda) $ is a failure rate that can be easily determined by, say, Monte Carlo methods. Then, we evaluate ratios $R_j$ using the acceptance ratio method due to Bennett~\cite{Bennett76}. The acceptance ratio method expresses $R_j$ as the fraction of two expectation values that can be evaluated efficiently via the Metropolis-Hastings algorithm. Details of the algorithm and its implementation are given in App.~\ref{App:Splitting}.

\begin{figure}
\includegraphics[width=\columnwidth]{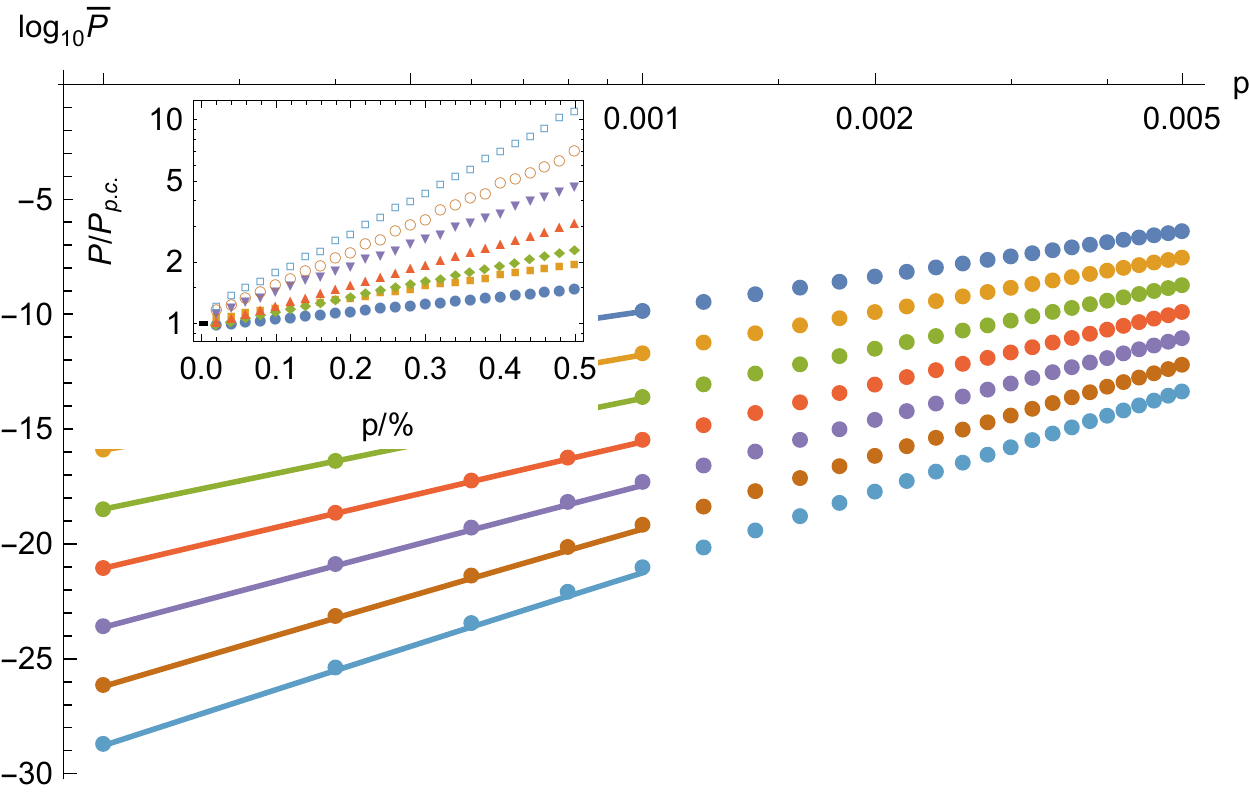}
\caption{\label{Fig:WenPathCounting} Logical failure rates obtained using the numerical method due to Bravyi and Vargo~\cite{Bravyi13} compared with the low physical error rate bound given in Eqn.~(\ref{eqn:pcW}) as a function of physical error rate $p$ for system sizes $\sqrt{n} = 10,\,12,\dots,\,22$. The inset shows the ratio of the failure rates that we obtained numerically and the low error rate bound. The convergence to unity as $p$ vanishes shows good agreement with the approximation.}
\end{figure}

In Fig.~\ref{Fig:WenPathCounting} we show the logical failure rates for the rotated model at system sizes $\sqrt{n} = 10,\,12,\dots,\,22$ to error rates as low as $p = 2 \cdot 10^{-4}$. As we explain in detail in App.~\ref{App:Splitting}, we evaluate each expectation value using the Metropolis-Hastings algorithm where we propose $N = 10^9 $ new trials for each expectation value, and find that at least $ \sim 5 \cdot 10 ^5 $ different error configurations are accepted for each calculation for distributions at the lowest $p$ values we investigate. For larger values of $p$ we find that as many as $\sim  10^8 $ different error configurations of the $10^9$ configurations that are proposed during the Metropolis Hastings algorithm are accepted for a given expectation value.

We use the method to compare the data with the low error rate limit we obtained in Eqn.~(\ref{eqn:pcW}). Our results show that the data converges onto our limit, thus verifying the predictions made in the previous section. To better illustrate the convergence of the data, the inset of Fig.~\ref{Fig:WenPathCounting} shows the ratio of the numerically evaluated failure rates and the analytical expression. The inset shows the ratio $\overline{P}^\W / \overline{P}^\W_{\text{low-}p}$ approaches $1$ from above as the physical error rate vanishes. This behaviour is to be expect as higher order terms in the logical error rate become less appreciable as $p$ decreases.

Together with an explanation of the method due to Bravyi and Vargo, we provide an equivalent analysis using the square-lattice model in App.~\ref{App:Splitting}. The comparison with the well-established expression for logical failure rate demonstrates the accuracy of the method at low error rates.

\begin{figure}
\includegraphics[width=\columnwidth]{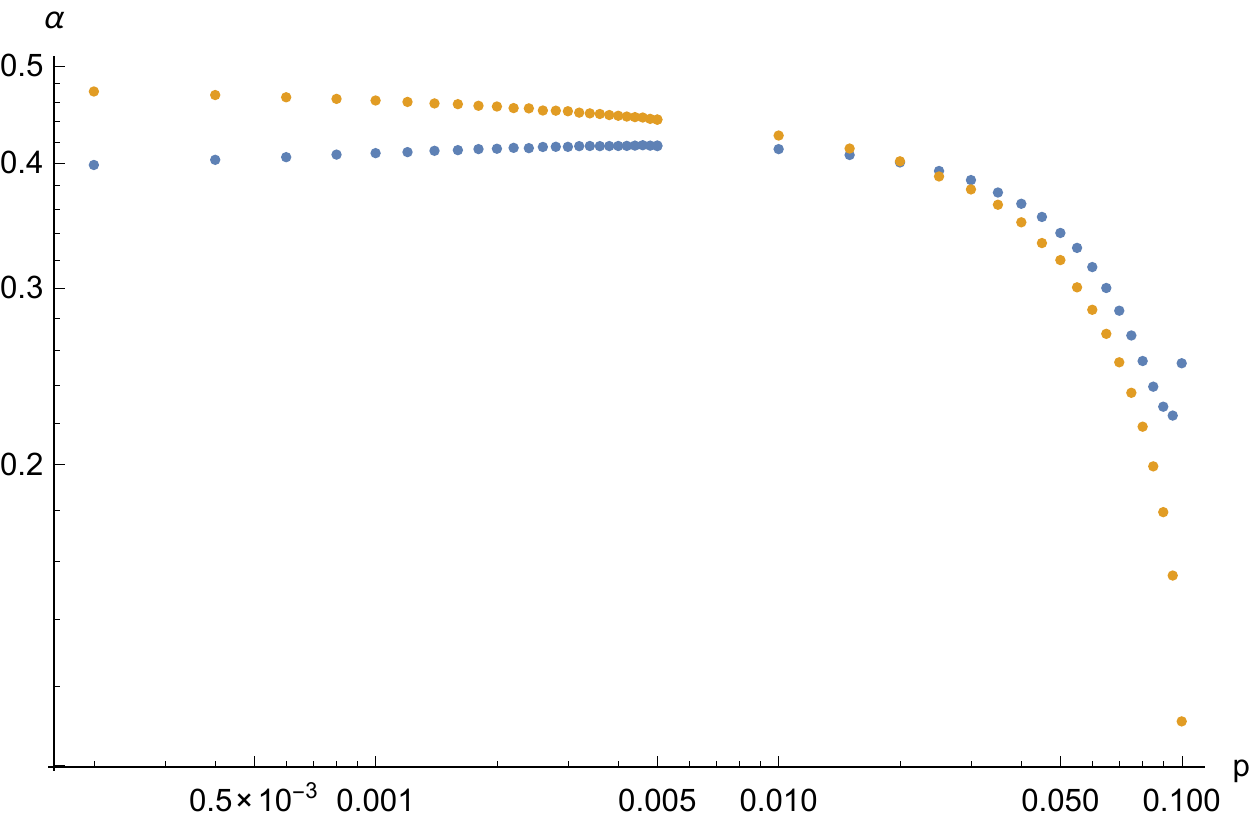}
\caption{\label{Fig:alpha} The function $\alpha(p)$ from the fitting Ansatz using Monte Carlo samples for $p > 0.05$ and the splitting method otherwise. We collect data for system sizes $10\le n\le 22$ as shown in Figs.~\ref{Fig:KitSplitData} and~\ref{Fig:WenSplitData}. The square(rotated)-lattice model is shown in blue(yellow). 
We observe slow convergence of $\alpha$ to $1/2^{3/2}$($1/2$) for the square(rotated)-lattice models as we have predicted using the path-counting formulae presented in the previous Section. We also observe a crossing in the functions $\alpha^\K(p)$ and $\alpha^\W(p)$ at around $p \sim 2\%$. }
\end{figure}

We finally examine more closely the logical failure rates as the error rate vanishes. In Fig.~\ref{Fig:alpha} we plot the fitting function $\alpha(p)$ of Eqn. (\ref{Eqn:MCFittingAnsatz}) for the two models for system sizes in the interval $10 \lesssim \sqrt{n} \lesssim 22$ over an extensive range of $p$. The logical failure rates for intermediate error rates for the square-lattice and rotated-lattice models that have not been presented explicitly so far in the text are shown in Figs.~\ref{Fig:KitSplitData} and~\ref{Fig:WenSplitData}, respectively in Appendix.~\ref{App:Splitting}. The plot shows that $\alpha^\K(p) $ tends towards $1/ 2\sqrt{2}$ and $\alpha^\W(p)$ tends towards $1/2$ as $p$ vanishes, as expected from the low error rate analysis given previously.

We observe that the values of $\alpha(p)$ for the two models cross at around $ p \sim 2 \%$ in Fig.~\ref{Fig:alpha} such that the square-lattice model will outperform the rotated lattice model if we extrapolate the system size. This is in contrast to conclusion obtained with the Monte-Carlo analysis where we study larger system sizes, see Fig.~\ref{Fig:LargeAlpha}. Indeed, we find that, in fact, $\alpha(p)$ varies slowly with $n$. As such, the Ansatz given in Eqn.~(\ref{Eqn:MCFittingAnsatz}) needs to be modified to account for the $n$ dependence of the functions $A$ and $\alpha$. Here we have implicitly assumed that the drift in these values is slow in $n$ such that we can use a linear fit for $\log_{10}\overline{P}$ in $\sqrt{n}$, provided we only study a small interval of $n$, see Figs.~\ref{Fig:KitSplitData} and~\ref{Fig:WenSplitData}.

\begin{figure}
\includegraphics[width=\columnwidth]{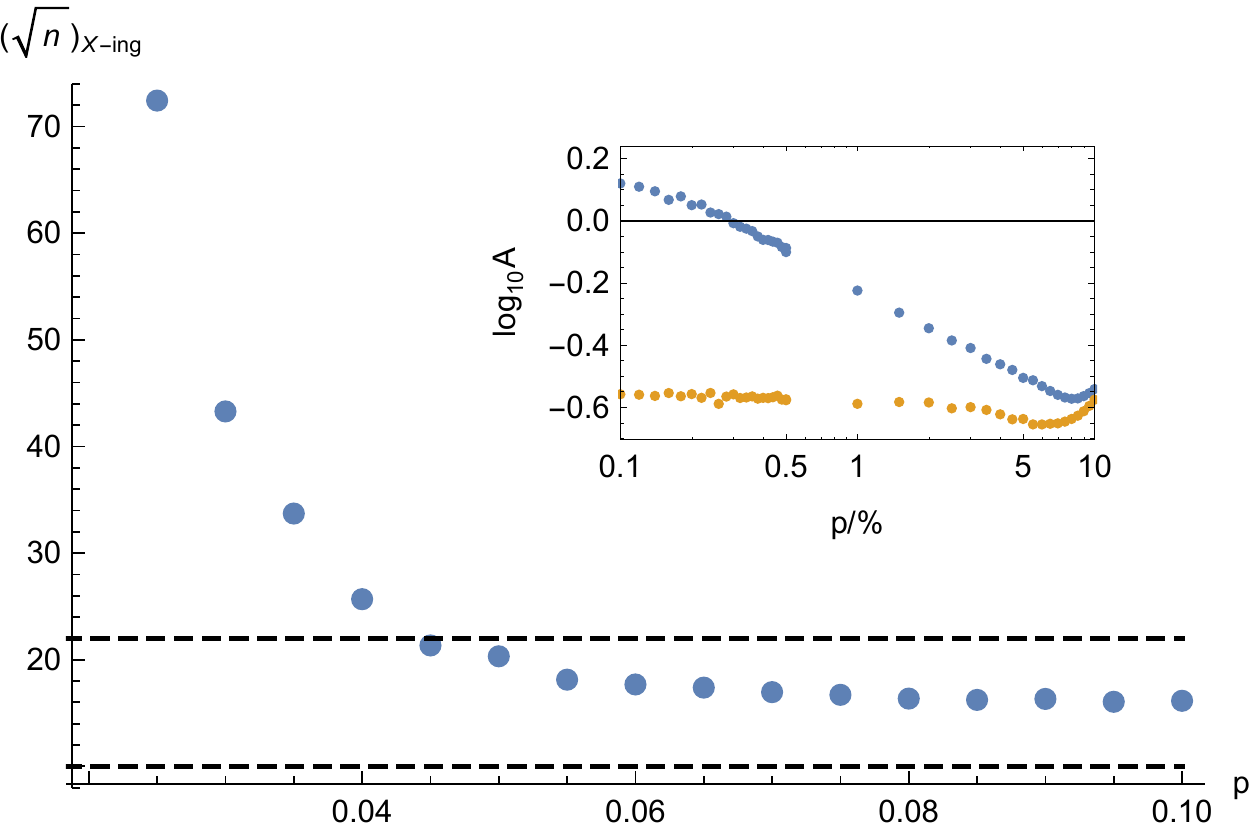}
\caption{\label{Fig:logA} The system sizes at which the square-lattice model begins to outperform the rotated-lattice model as $n$ increases. We also mark $\sqrt{n} = 10$ and $\sqrt{n} = 22$ by black dashed lines to indicate where the range of system sizes for which we collect data. These data points mark the lower boundary of the hatched region shown in Fig.~\ref{fig:summaryintro}. The plot of $\log A$ shown as a function of $p$ between the path-counting regime and the threshold error rate for the square-lattice model(blue) and the rotated-lattice model(yellow) is shown in the inset. } 
\end{figure}

We finally compare the logical failure rates of the two models within the smaller system sizes where $10 \lesssim \sqrt{n} \lesssim 22$. Indeed, due to discrepancies in $\log_{10}A(p)$, we find that for system sizes $\sqrt{n} \lesssim 16$, the rotated-lattice model outperforms the square lattice model. We show the values of $\log_{10}A(p)$ in the inset of Fig.~\ref{Fig:logA}, where these values are evaluated using the intersect of the fitting with $n = 0$ found in Figs.~\ref{Fig:KitSplitData} and~\ref{Fig:WenSplitData}. We mark the points where the linear fittings for the square-lattice model and the rotated-lattice model cross in the main plot in Fig.~\ref{Fig:logA}. These data points mark the bottom boundary of the hatched region, thus completing our numerical estimate of the location of the boundary of the parameter regime where the square-lattice model outperforms the rotated model. It will be interesting to explain the change in $\alpha(p)$ as $n$ increases throughout the hatched region.

\section{Modeling finite physical error rates}\label{sec:model}

In Sec. \ref{sec:pathcounting}, we considered the first-order analytical approximation for asymptotically low error rate regime for arbitrarily large system size. 
Here we generalise path counting methods to model topological quantum error correction at intermediate error rates. We will argue that the following expression \begin{equation}
\overline{P}_{\text{model}} = \sum_{l=d}^{n} N_{\text{con}}(l) ~\xi(p)^l~ p^{\frac{l}{2}} \left( 1 -p \right)^{\frac{l}{2}}, 
\label{eq:generalclosedmodel}
\end{equation}
provides a qualitatively accurate  expression for the failure rate, where $N_{\text{con}}(l)$ is the number of non-self intersecting paths of length $l$ that wrap around the torus.  
The model can be interpreted as a statistical mechanics model of a string wrapping around the torus in the presence of a `background gas'. 
The $\xi(p)$ term in Eqn.~(\ref{eq:generalclosedmodel}) describes the interaction which can be seen as a `negative friction' term  encouraging the string to fluctuate.
%Our model is no longer exact, and relies on a number of reasonable approximations.

\subsection{Upper bounding the failure probability, and recovering a lower bound of the threshold}

The following discussion applies to either orientation of the surface code on $n$ qubits.
As described in Section~\ref{sec:entropy}, the probability of a logical error is
\begin{eqnarray}
\overline{P}(p,n) &=& (1-p)^{n} \sum_{w=d/2}^{n} N_{\text{fail}}(w) \left(\frac{p}{1-p}\right)^{\!\! w},
\end{eqnarray}
where $N_{\text{fail}}(w)$ is the number of weight-$w$ elements of the failing error set $\mathcal{F}$. 
To obtain a more transparent expression for $\overline{P}$, we need to characterize which bit strings are contained in $\mathcal{F}$. 
The product of an error operator and its subsequent correction, $C(E)E$, is supported on closed paths on the lattice in the form of stabilizer operators or of logical operators.
When a failure occurs, at least one of the closed paths in $C(E)E$ must be non-contractible -- we call the subset of edges that form this single non-contractible closed path $L$.

Suppose that $C(E)E$ contains some specific non-contractible closed path $L$ such that a logical error occurs. 
It must be that $L \cap E$ has weight greater than $|L|/2$, since otherwise the minimum weight decoder would have yielded a lower weight correction $C'(E) = C(E) L$, where $L$ is absent in $C'(E)E$.
We can use this to help construct the following bound,
\begin{equation}\label{eqn:bound}
\! \overline{P} \leq (1-p)^{n} \! \sum_{L} \! \sum_{u=\frac{|L|}{2}}^{|L|} \! \sum_{v=0}^{n-|L|} \! C_{u}^{|L|}  C^{n-|L|}_{v} \left(\frac{p}{1-p}\right)^{\!\! u+v}. \!\!
\end{equation}
The outer sum is over all non-contractible, self-avoiding closed paths in the lattice.
Given such a closed path $L$, the inner sums add up the probability that an error occurs which has support on more than half of the edges of $L$.
To do this, we divide the lattice into the $|L|$ edges in the non-contractible closed path $L$ and the $n-|L|$ edges in its compliment. 
There are $C^{|L|}_{u}$ choices of $u$ edges along the closed path, and $C^{n-|L|}_{v}$ choices of $v$ edges outside it.
The probability of any error configuration with a total $u+v$ errors is $p^{u+v}(1-p)^{n-u-v}$.

It is useful to define $N_{\text{con}}(l)$, the number of length-$l$ non-contractible closed paths in the lattice, constrained by the requirement that they can have no self-intersections.
We then rewrite the bound as
\begin{eqnarray}
\overline{P} &\leq& (1-p)^{n} \sum_{l=d}^{n} N_{\text{con}}(l) \sum_{u=\frac{l}{2}}^{l} \sum_{v=0}^{n-l} C^{l}_{u} C^{n-l}_{v} \left(\frac{p}{1-p}\right)^{\!\! u + v}, \nonumber \\
& \leq & \sum_{l=d}^{n} N_{\text{con}}(l)~ 2^l ~p^{\frac{l}{2}} \left( 1 -p \right)^{\frac{l}{2}}.
\label{eq:bound}
\end{eqnarray}
We have used an explicit computation of the sum over $v$,
\begin{equation}
\sum_{v=0}^{n-l} C^{n-l}_{v}\left(\frac{p}{1-p}\right)^{\!\! v} =  \left( 1  -p \right)^{-n+l},
\end{equation}
and we also used the following bound for the sum over $u$,
\begin{equation}
\sum_{u=\frac{l}{2}}^{l}C^{l}_{u}  \left(\frac{p}{1-p}\right)^{\!\! u}  \leq 2^l  \left(\frac{p}{1-p}\right)^{\!\! l/2},
\label{eq:ineq}
\end{equation}
which holds for all $l$ and $0<p<1/2$.

Our formula in Eqn.~(\ref{eq:bound}) for an upper bound on the failure probability can be used to give a lower bound on the error correction threshold similar to that in \cite{Dennis02}. 
First, we use the bound that $N_{\text{con}}(l) < N_0 c^l$ (for a constant $N_0$ and with $c = 2.638\dots$). $c$ is the expansion constant for self-avoiding walks on the square lattice \cite{Madras96}.
Then we use Stirling's bound to group terms that are exponential in $l$.
Finally, we find the maximum $p$ below which the upper bound of $\overline{P}$ approaches zero for large $n$.
\begin{eqnarray}
\overline{P} &\leq& \sum_{l=d}^{n} N_{\text{con}}(l,n)~2^l~ p^{\frac{l}{2}} \left( 1 -p \right)^{\frac{l}{2}}, \nonumber \\
&\leq& \sum_{l=d}^{n} N_0 \cdot  \left( 2 c \sqrt{p(1-p)} \right)^{\! l}.~~~~~\label{eq:DennisBound}
\end{eqnarray}
For sufficiently large $n$, the right-hand side will approach zero with $p\rightarrow 0$ provided that the parenthesis is less than 1 in Eqn.~(\ref{eq:DennisBound}).
This gives a lower bound on the threshold of $p_{\text{th}}> p_{\text{bound}}=0.0373$, which is the same as the bound in Ref.~\cite{Dennis02}.

The sum in Eqn.~(\ref{eq:bound}) over-counts in two ways: some error configurations which are not present in $\mathcal{F}$ are included, and some error configurations are included more than once. 
Indeed, some distributions of $u> l/2$ in Eqn.~(\ref{eq:bound}) can be associated to several strings, and these are counted several times. 
%\mk{We can interpret the model as a statistical mechanics model of a string with fixed endpoints  in the presence of a 'background gas'. Eqn.~(\ref{eq:bound}) describes the non-interacting model, yet the exact expression involves an interaction term between the lattice gas and the string. The interaction can be seen as a 'negative friction' term that encourages the string to fluctuate. } {\color{blue} BJB - $\leftarrow$ is this analogy actually helpful? Do we use it later on? I suspect that we are hanging onto this because it was a cute way for us to understand as we built up this work, but maybe it needn't make the final draft.}
Our goal in the following subsections will be to remove terms from the sum, and to estimate $N_{\text{con}}(l,n)$ to get closer to the true value of $\overline{P}$. 
In doing so, we will lose the guarantee that we overestimate $\overline{P}$.

\subsection{Counting closed paths}
\label{sec:countingNonContractible}

\begin{figure}
	\includegraphics[width=\columnwidth]{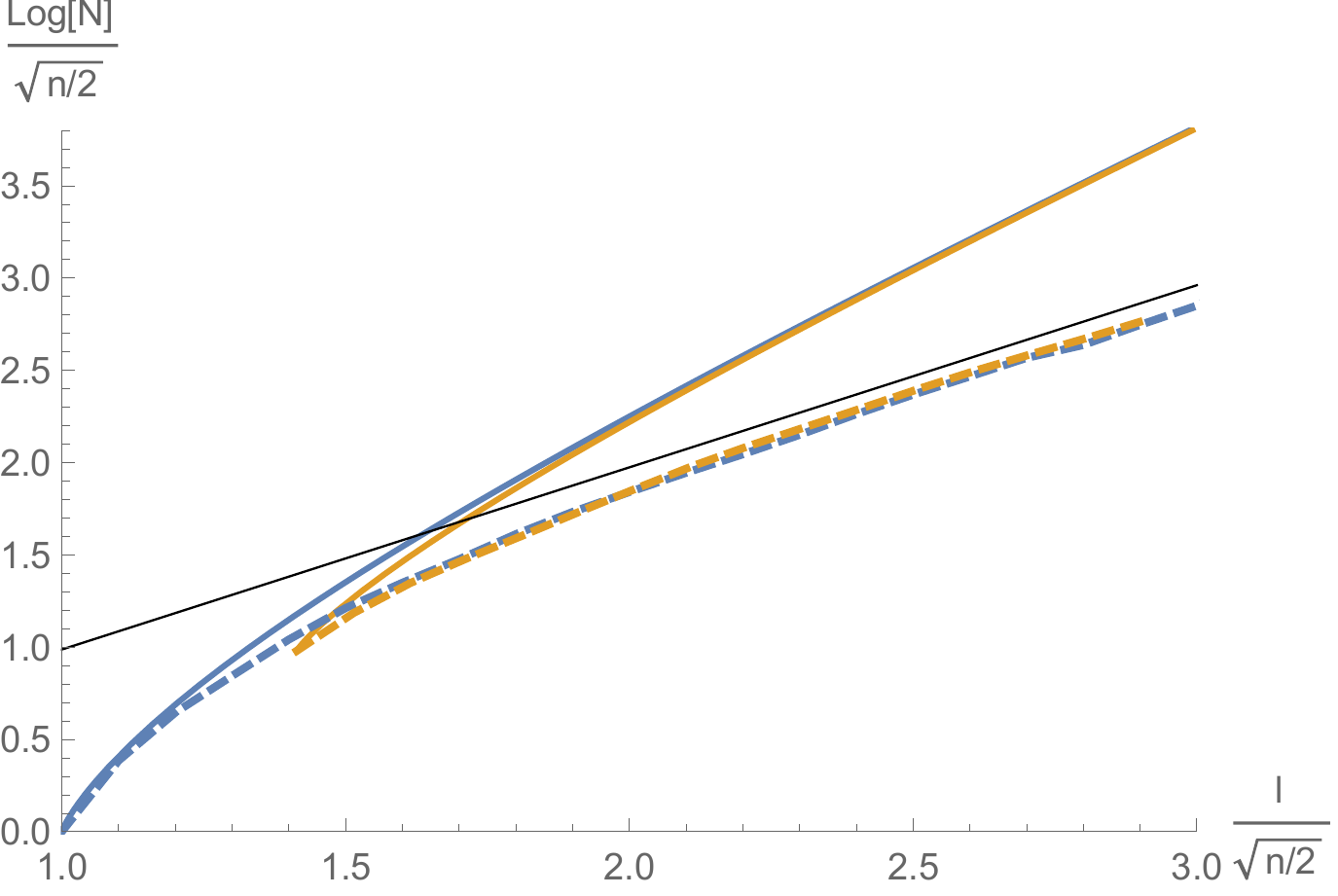}
	\caption{
		The logarithm of the number of paths, normalized by $\sqrt{n/2}$ in the limit of large $n$.
		The dashed curves represent our estimates of $N^{\K}_{\text{con}}(l,n)$ and $N^{\W}_{\text{con}}(l,n)$, and the solid curves are the exact limit of $N_{\text{unc}}(l;x,y)$ for $(x,y)=(\sqrt{n/2},0)$ and $(x,y)=(\sqrt{n}/2,\sqrt{n}/2)$ as calculated using Eq.~(\ref{eq:gaussianintegral}).
		Note that the two curves asymptotically approach one another for large $l/\sqrt{n/2}$.
		The blue and yellow curves are for the square diamond lattice respectively.
		The number of constrained and unconstrained paths match for $l=d$ for both square and diamond lattice orientations.
		The black line which both $N^{\K}_{\text{con}}(l,n)$ and $N^{\W}_{\text{con}}(l,n)$ appear to be approaching is $N=c^l$ where $c= 2.638\dots$ is the square-lattice connective constant.}
	\label{fig:PlotUnconstrainedPaths}
\end{figure}

An important quantity in Eqn.~(\ref{eq:bound}) is $N_{\text{con}}(l,n)$, the number of length-$l$, non-contractible, self-avoiding closed paths on the lattice.
In this section we estimate $\log[N_{\text{con}}(l,n)]$ in the limit of large $n$ as shown in Fig.~\ref{fig:PlotUnconstrainedPaths}.

\subsubsection{Number of unconstrained paths}
\label{subsec:unconstrained}

We can gain  insight into the behaviour of $N_{\text{con}}(l,n)$ by considering a closely related quantity, the number of unconstrained paths $N_{\text{unc}}(l;x,y)$, which is much easier to calculate. 
We define $N_{\text{unc}}(l;x,y)$ to be the number of length-$l$ paths from the origin to the coordinate $(x,y)$ on an infinite square lattice, where the axes are aligned along the edges of the respective lattices, see Fig.~\ref{Fig:Lattices}.

When $l=x+y=d$, the path is perfectly tight, and the function $N(l;x,y)$ can be precisely related to $N_{\text{con}}(l,n)$ 
\begin{eqnarray}
N_{\text{con}}^{\K}(d,n) & = & 2 \cdot d \cdot N_{\text{unc}}(d;d,0),\label{eqn:NconNuncK}\\
N_{\text{con}}^{\W}(d,n) & = & 2 \cdot \frac{d}{2} \cdot N_{\text{unc}}(d;d/2,d/2) + d.\label{eqn:NconNuncW}
\end{eqnarray}
To see why these relations hold, first note that $N_{\text{unc}}(d;d,0)=1$, i.e. there is just one length-$d$ path from the origin to the point $(d,0)$.
The multiplicative factor of $2$ accounts for the fact that  in addition to the horizontal cycle on the torus, there is also a vertical cycle.
The factor of $d$ accounts for translations;  the horizontal path could go through the point $(0,j)$ for $j=1,2,\dots d-1$ rather than $(0,0)$.
Similarly, the factor of $d/2$ for the rotated lattice accounts for the fact that the horizontal paths could go through the point $(-j,j)$ for $j=1,2,\dots d/2$ rather than $(0,0)$.
The additional constant contribution $d$ for the rotated lattice counts the non-contractible closed paths which wind simultaneously both horizontally and vertically around the torus.
For $l$ larger than $d$, we do not know of any clean relation like Eqn.~(\ref{eqn:NconNuncK}) and Eqn.~(\ref{eqn:NconNuncW}) between the number of constrained and unconstrained paths. 
%{\color{blue} I might suggest relating this last paragraph back to section III. - BJB}

We assume in the following that $l$ is even. To calculate $N_{\text{unc}}(l;x,y)$ we assign an orientation; up, down, left or right, denoted $\{\uparrow, \downarrow, \leftarrow, \rightarrow\}$, to each of the edges in the sequence of $l$ contiguous edges on a path from the origin to the point $(x,y)$. The direction of each edge is determined by the direction the path follows through the edge towards its terminal point. Obviously, the difference between the number of right edges and the number of left edges must equal $x$, i.e. $n_\rightarrow - n_\leftarrow = x$. Likewise, the difference between the number of up edges and down edges must equal $y$. Again, expressed as an equation we have $n_\uparrow - n_\downarrow = y$. By definition, we also have that $n_\uparrow + n_\downarrow +n_\rightarrow + n_\leftarrow = l$.
Using these three conditions, we can rewrite $n_\downarrow, n_\rightarrow$ and  $n_\leftarrow$ in terms of $n_\uparrow$ such that
\begin{equation}
N_{\text{unc}}(l;x,y)= \!\!\!\sum_{n_\uparrow = y}^{(l + y - x)/2} \!\!\! \frac{l !}{n_\uparrow!  n_\downarrow(n_\uparrow)!  n_\rightarrow(n_\uparrow)!  n_\leftarrow(n_\uparrow)!}. \!\!
\label{eq:Nunconstrained}
\end{equation}
In Appendix~\ref{app:unconstrainedpaths} we find a closed form expression for $\log[N_{\text{unc}}(l;x,y)]$, which we plot in Figure~\ref{fig:PlotUnconstrainedPaths}. 
Defining $r$ and $\theta$ by $x=r \cos\theta$ and $y=r \sin\theta$ expanding the expression in powers of $r/l$ yields
\begin{eqnarray}
\frac{\log[N_{\text{unc}}]}{l} &\! = \! &  \log[4] - \frac{r^2}{l^2} \!+\! \frac{(\cos[4 \theta] \!-\! 3)r^4}{12 l^4} \!+\! \mathcal{O}\left(\frac{r^6}{l^6}\right)\!.~~~~
\end{eqnarray}
This is quite informative as it suggests that for loose strings where $l \gg r$ the two lattice orientations have the same $N_{\text{unc}}(l)$.
We know that for tight strings, for which $l \approx x+y$, the behaviour of $N_{\text{unc}}(l)$ is very different for the two lattice orientations. This phenomenon can be regarded as a way of recovering the Euclidean distance from the Manhattan distance. Indeed, the degeneracy of the paths to reach one point from another restores a Euclidean like metric over a square lattice that would otherwise respect the square lattice geometry. We suggest that this feature is responsible for our numerical results that show that the logical error rate for both orientations are similar at threshold.

\subsubsection{Estimating the number of constrained paths}
\label{subsec:constrained}

Although we cannot calculate $N_{\text{con}}(l,n)$ exactly in the large $n$ limit, we can estimate it in two stages. 
First, we estimate $N_{\text{con}}(l,n)$ for a sequence of system sizes by randomly sampling  unconstrained paths that contribute to $N_{\text{unc}}(l;x,y)$ and counting what fraction happen to satisfy the non-self intersection property. 
Second, we extrapolate $N_{\text{unc}}(l;x,y)$ by fitting to a functional form that allows us to estimate the limit for asymptotically large $n$.
We provide further details in Appendix~\ref{sec:FiniteSizeNcl}.
The results are shown for the square and diamond lattice in Fig.~\ref{fig:PlotUnconstrainedPaths}.
We summarize some key features. 
\begin{enumerate}[(i)]
\item $N^{\K}_{\text{con}}(l,n)=0$ for $l<\sqrt{n/2}$ and $N^{\W}_{\text{con}}(l,n)=0$ for $l<\sqrt{n}$, implying that if the term multiplying $N_{\text{con}}(l,n)$ in Eqn. (\ref{eqn:bound}) is strongly suppressed when $l>\sqrt{n}$, then the failure rates will be very different. 
\item Both $N^{\K}_{\text{con}}(l,n)$ and $N^{\W}_{\text{con}}(l,n)$ approach $c^l$ in the limit $l/\sqrt{n/2} \rightarrow \infty$ from numerical studies \cite{Madras96}, where $c\approx 2.638$ is the expansion constant.
	This should be contrasted with the unconstrained case in which $N^{\K}_{\text{unc}}(l;x,y)\sim 4^l$. The asymptotic scaling also appears to be reached more rapidly for the unconstrained paths than for the constrained ones. 
	\item  $N^{\W}_{\text{con}}(l,n)<N^{\K}_{\text{con}}(l,n)$ for $l=\sqrt{n}$, and although our estimates indicate that the two functions cross, it is hard to conclusively claim this with our data.
\end{enumerate}

\subsection{Estimating the failure probability}

Our goal here is to justify and further analyse the model of Eqn. (\ref{eq:generalclosedmodel}). 
Note that the rigorous upper bound in Eqn. (\ref{eq:DennisBound}) dramatically over counts the failure modes, since many of the $C^{l}_{l/2}\sim 2^l$ configurations of errors along a given path are counted multiple times as they are also found as configurations on other paths. 
In Eqn. (\ref{eq:generalclosedmodel}), we have corrected for this by replacing $2^l$  by $\xi(p)^l$.
This is no longer a rigorous bound, but forms a relatively simple model applicable for all $p$ and arbitrarily large $n$.
There are two limits of $1 \leq \xi(p) \leq 2$. 
The limit $\xi(p) = 2$ implies that any of the $ C^{l}_{l/2} \sim 2^l$ configurations of errors along a particular string of length-$l$ will result in a correction to that specific string.
On the other hand, the limit $\xi(p) = 1$ implies that every string of length-$l$ is associated with just a single configuration of errors along the string. %\mjk{Should somehow promote Eqn. (\ref{eq:generalclosedmodel}). This is after all our NEW model. } {\color{blue} BJB - agree. I'll leave this to someone else, but I suggest promoting the equation to the top of the paragraph, and justifying it afterwards.}
%{\color{magenta} MB - I guess we could write it at the very start of section V before V A starts and say that is where we are headed, and explain what Ncon is roughly?.}

The intuition for the introduction of the phenomenological term $ \xi(p)$ comes from our analysis of the low error rate regime in Section~\ref{sec:pathcounting}.
In that regime, all errors of a failing configuration lie along the non-contractible closed path which results after correction, and we were able to identify precisely which configurations of $d/2$ errors along a particular length-$d$ path would correct to that path. 
We found that all configurations along a completely straight path resulted in a correction to include that path, whereas only configurations with errors in some of the corners of curved paths would do so. 
From this perspective, it would make sense to include an explicit dependence of $\xi$ on $l/d$, since larger $l/d$ corresponds to paths with more curves.
However, for a given $p$, the main contributions to the sum in Eqn.~(\ref{eq:generalclosedmodel}) are from a narrow range of $l/d$, such that $p$ is essentially a proxy for $l/d$.

Away from the low error limit, there are an extensive number of errors that do not lie on the string as well, that inevitably 'interact' with the string.
Below the threshold, these errors should generically not prevent a configuration from failing, although they can result in an over counting of a failing configuration, since more than one path of the same length $l$ could have $l/2$ errors in it.

\subsubsection{Identifying $\xi(p)$}

\begin{figure}
	\includegraphics[width=\columnwidth]{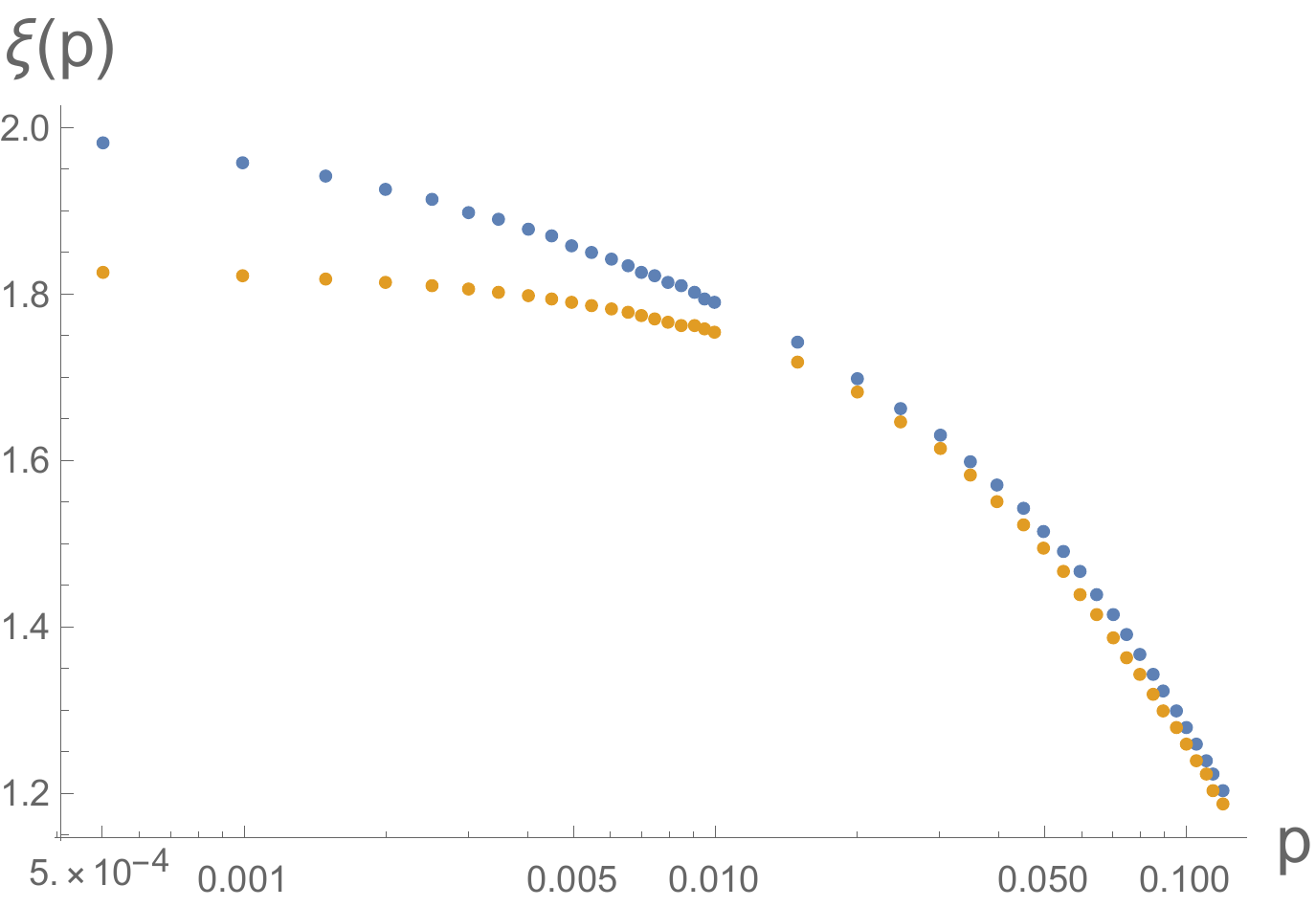}
	\caption{We identify $\xi(p)$ numerically by fitting the data obtained by Monte Carlo and the splitting method across a large range of $p$ to the model in Eq.~(\ref{eq:generalclosedmodel}).
		We used $d=10$ for the rotated lattice, and $d=14$ for the square lattice. 
		The $N_{ncl}(l)$ were estimated to within 2\% accuracy for both models by sampling, as described in Appendix~\ref{sec:FiniteSizeNcl}.
		There are some deviations from the expected behaviour, which we expect comes from small-size effects. 
		For example, the blue curve appears to be slightly below the expected value of 2 for $\xi^{\K}(0)$, and the values $\xi^{\K}(p_{\text{th}})$ and $\xi^{\W}(p_{\text{th}})$ appear to differ slightly.
	}
	\label{fig:XiPlot}
\end{figure}

By fitting the model in Eqn.~(\ref{eq:generalclosedmodel}) to the data for some small system sizes, we provide a numerical estimate of $\xi^{\K}(p)$ and $\xi^{\W}(p)$in Fig.~\ref{fig:XiPlot}. We observe a change of behavior around $p\sim 1\%$, above which both orientations have near identical values of $\xi(p)$ and below which they diverge. This points to the fact that for low error rates, the free energy at fixed energy has a higher entropy contribution for the original lattice over the rotated lattice for the entire range of sting lengths; since both contributions $\xi(p)$ and $N_{\rm con}(l)$ are larger for the original lattice.

For completeness, we estimate the value of $\xi(p)$ in the limits of $p\rightarrow 0$ and $p\rightarrow p_{\rm th}$.
Keeping only the lowest order terms in $p$, and comparing with the large $n$ expressions from Section~\ref{sec:pathcounting},
\begin{eqnarray}
\overline{P}_{\text{model}} & \rightarrow & N_{\text{con}}(d) ~\xi(0)^d~ p^{\frac{d}{2}}.\\
\overline{P}_{\text{low-p}} & = &  (\gamma_0)^{\sqrt{n}}~p^{\frac{d}{2}}.
\end{eqnarray}
Comparing these, and given $N_{\text{con}}^{\K}(d) = 2 \sqrt{n/2} \sim 1^{\sqrt{n}}$, $N_{\text{con}}^{\W}(d) = d C^{d}_{d/2} + d \sim 2^{\sqrt{n}}$ and assuming that $\gamma^\K = \sqrt{27/2}$ matches the upper bound in Eq.~(\ref{eqn:tighter}), we identify
\begin{eqnarray}
\xi^{\K}(0) & = & 2, \\
\xi^{\W}(0) & = & \sqrt{27/8} \approx 1.8371.
\end{eqnarray}
These values are encouragingly close to the values extracted from the model in Fig.~\ref{fig:XiPlot}.

%We will now argue on heuristic grounds that $O(|\xi(0)-\xi(\epsilon)|)=\epsilon$ for small enough, but constant, $\epsilon$. In the low error rate regime, all of the errors are sprinkled somewhere on the error line; i.e. there are $O(\sqrt{n})$ errors. 
%For $p\ll 1$ but constant, there is an extensive $p n \propto n$ errors. 
%All but a fraction of $O(1/\sqrt{n})$ of the extra errors will contribute to the background gas, and not affect the string length or the number of strings. 
%However, all new errors within a band of width roughly $\epsilon \sqrt{n}$ might interact with the string, changing its entropic contribution.  However, this contribution will be at most of order $ C^{l}_{\epsilon l}$, hence contributing a modification of order $\epsilon$ to $\xi(\epsilon)$.

Near threshold, we expect that the dominant contributions to Eqn.~(\ref{eq:generalclosedmodel}) are from terms with  $l\gg d$. 
Therefore,  a typical string will be loose (i.e. have a lot of kinks), and  there are $\sim p n$ ambient errors not on the string -- which implies that many of the configurations of  $ C^{l}_{l/2} \sim e^{l\log[2]}$ errors along a given string will \textit{not} correct to give that particular string. 
This implies that near threshold, we should expect $\xi(p_{\text{th}})$ to be close to $1$; this would correspond to a single string per error configuration. %, which is the large $p$ limit. 
Near the threshold, we also suppose that the sum is dominated by terms for which the string length is in the regime $N_{\text{con}}(l,n) \sim c^l$ for $c \approx2.638$ for both lattice orientations. 
Furthermore, since $N_{\text{con}}(l,n)$ is equal for both models at threshold and the threshold is the same for both orientation (since the threshold is a bulk property), then $ \xi^{\K}(p_{\text{th}}) = \xi^{\W}(p_{\text{th}}) \equiv \xi_{\text{th}}$ since the expression in Eqn.~(\ref{eq:generalclosedmodel}) are otherwise the same for both orientations
\begin{eqnarray}
\overline{P}_{\text{model}} & \rightarrow & \sum_{l=d}^{n} \left(c ~\xi_{\text{th}} \sqrt{p \left( 1 -p \right)} \right)^{l}.
\end{eqnarray}
Let $p_c$ be the value of $p$ at which the parenthesis becomes one.
For $p<p_c$, the parenthesis is less than one, such that the sum is dominated by small $l$, for which the assumption that $N_{\text{con}}(l,n) \sim c^l$ would break down.
For $p>p_c$ on the other hand, the exponent is maximized for large $l$ and the total failure probability will become large.

Solving for $p_c$, we obtain
\begin{eqnarray}
p_c=\frac{1}{2} - \sqrt{\frac{1}{4}-\frac{1}{\xi_{\text{th}}^2c^2}}.
\end{eqnarray}
%When $\xi_{\text{th}} = \log[2]$, $p_c=17.40 \%$.
The actual threshold value is known to be $p_{th} = 10.3\%$, which corresponds to $\xi_{\text{th}} \approx 1.2471$. 
This is close to the value obtained from the data in Fig.~\ref{fig:XiPlot}.

\section{Conclusions}

We have taken a number of different approaches to explore the configuration space of errors that can cause logical failure for the surface code with different lattice geometries. We have found that, while it pays to optimize the distance of the code, entropic factors can have a significant effect on the performance of codes at modest physical error rates.

It will be interesting to explore the entropic contribution on other codes with improved encoding rates such as twisted surface codes~\cite{Yoder16}, color codes~\cite{Bombin06, Landahl11}, stellated color codes~\cite{Kesselring18} and hyperbolic codes~\cite{Freedman01, Freedman02, Delfosse13, Breuckmann16, Breuckmann17} to determine how the entropic considerations affect logical failure rates here. Indeed, one could imagine that codes that require a reduced  number of physical qubits to realize a code of a given distance may suffer adversarially from entropic effects. Further, for the system sizes we have studied, the logical failure rate of the two different codes are almost indistinguishable until the physical error rate is an order of magnitude below threshold. We might like to account for this when we consider the change in distance of fault-tolerant quantum systems as we perform logical operations~\cite{Brown16a}.

It may also be worthwhile studying entropic effects during fault-tolerant error correction. Correlated errors that occur during syndrome extraction manifest themselves as diagonal bonds during error correction~\cite{Fowler09, Fowler12}. Extensions of the present work may consider choosing circuits to minimize these effects. Recently, flag fault tolerance~\cite{Chao18, Chao17} has been considered in topological codes to minimize these correlated errors~\cite{Chamberland18}. One might view the extra resources used to implement these ciruits as an additional hardware expense used to reduce logical failure rates by minimzing the configuration space of errors.

Ultimately, it will be very useful to determine bounds on the extent to which the physics of quantum error-correcting codes will permit us to minimize entropic factors in the logical failure rate to help us to design better fault-tolerant quantum-computational protocols in the future.

\begin{acknowledgements}
We are grateful for helpful and supportive discussions with H. Bomb\'{i}n, N. Delfosse, S. Flammia, R. Harper, A Hutter, D. Poulin, J. Preskill, S. Simon and J. Wootton. The authors acknowledge the facilities, and the scientific and technical assistance of the Sydney Informatics Hub at the University of Sydney and, in particular, access to the high performance computing facility Artemis. BJB and MJK are supported by Villum Fonden. MJK ackowledges support from the Deutsche Forschungsgemeinschaft (DFG, German Research Foundation) -- Projektnummer 277101999 -- TRR 183 (project B02). BJB is also supported by the University of Sydney Fellowship Programme and the Australian Research Council via the Centre of Excellence in Engineered Quantum Systems(EQUS) project number CE170100009.
\end{acknowledgements}

%\begin{widetext}

%%%%%%%%%%%%%%%%%%%%%    Appendix    %%%%%%%%%%%%%%%%%%%%%%%%

\appendix

%%%%%%%%%%%%%%%%%%%%%%%%%%%%%%%%%%%%%%%%%%%%

\section{Minimum weight decoding with low-weight errors}\label{app:decoder}

Here we make some general comments about minimum weight error correction decoders, which select the smallest correction given the syndrome.
These comments are particularly relevant for Sec.~\ref{sec:pathcounting} where we calculate bounds valid for the number of failure modes for all choices of minimum weight decoders.

A minimum-weight matching decoder applies a correction of minimum weight. 
However, the minimum-weight decoder is not fully defined by this rule alone, since there could be many corrections with the same minimum weight, and some of them could be logically inequivalent from one another. 

An optimal decoder selects the correction by identifying the most probable equivalence class of errors with the observed syndrome. 
In the limit of small $p$, the most probable equivalence class of errors is the one with the most minimum-weight errors. 

This allows us to define two minimum-weight decoders for a given code: The {\it best min-weight decoder} applies a correction from the largest equivalence class of weight $d/2$ errors consistent with the syndrome.
The {\it worst min-weight decoder} applies a correction from the smallest equivalence class of weight $d/2$ errors consistent with the syndrome.
In Table~\ref{tab:decodersWen}, we enumerate the number of minimum-weight errors which cause failures of different types (vertical or horizontal, or diagonal) for the rotated lattice.
In Table~\ref{tab:decodersKit} we show the same for the standard lattice.
We do not separate horizontal and vertical errors for the rotated code because there can be multiple error cosets of the same size, and choosing one rather than another can shift errors between horizontal and vertical failure modes. 
There are no minimum weight diagonal failures for the standard code.

\begin{table}[!htb]
	\centering
	\begin{tabular}{{c}l*{2}{c}}
		Distance & Decoder              & ~~~~ Horiz. or Vert.~~~~ & Diag. \\
		\hline
		\hline
		$4$ & Largest coset & 48 & 8  \\
		$4$ & Smallest coset & 48 & 8  \\
		\hline
		$6$ & Largest coset & 678  & 51   \\
		$6$ & Smallest coset & 822  & 51   \\
		\hline
		$8$ & Largest coset & 8752  & 264   \\
		$8$ & Smallest coset & 12144 & 264   \\	
	\end{tabular}
	\caption{The number of failing errors for the best and worst minimum weight matching decoders for the rotated code.}
	\label{tab:decodersWen}
\end{table}

\begin{table}[!htb]
	\centering
	\begin{tabular}{{c}l*{2}{c}}
		Distance & Decoder              & ~~~~ Horiz. ~~~~ & Vert. \\
		\hline
		\hline
		$4$ & Largest coset & 12 & 12  \\
		$4$ & Smallest coset & 12 & 12  \\
		\hline
		$6$ & Largest coset & 60  & 60   \\
		$6$ & Smallest coset & 60  & 60   \\
		\hline
		$8$ & Largest coset & 280  & 280   \\
		$8$ & Smallest coset & 280 & 280   \\	
	\end{tabular}
	\caption{The number of failing errors for the best and worst minimum weight matching decoders for the rotated code.}
	\label{tab:decodersKit}
\end{table}

Although there are efficient decoders that take into account the relative size of cosets \cite{Stace10, Duclos-Cianci10, Fowler13,Criger18,Wootton12, Bravyi14,Anwar14}, we do not attempt to do so in our numerical implementations.

In Sec.~\ref{sec:pathcounting} we found bounds for $N_{\text{fail}}(d/2)$ for each model, where $N_{\text{fail}}(d/2)$ is the sum of the number of horizontal, vertical and diagonal errors. 
For the standard model, $N_{\text{fail}}^{\K}(d/2)$ is independent of the choice of minimum weight decoder as is clear from Table~\ref{tab:decodersKit} (and is also straightforward to prove). 
On the other hand for the rotated model, from Table~\ref{tab:decodersWen} we can clearly see that $N_{\text{fail}}^{\W}(d/2)$ depends on which minimum weight matching decoder is used. 

In the rest of this appendix section we give an argument that the decoder we implemented numerically performs close to optimally.
First we argue that, since the minimum weight decoder implemented for our numerical analysis was chosen without regard to performance, we can presume that it performs similarly to choosing a correction at random from the set of corrections consistent with the syndrome, i.e. the {\it random minimum weight decoder}.

Second, we now prove that the random minimum weight decoder has the same $\gamma$ as the optimal minimum weight decoder. 
To see this, suppose there is a syndrome $\sigma$, consistent with four different error classes of size $M_{1}(\sigma) \geq M_{2}(\sigma) \geq M_{3}(\sigma) \geq M_{4}(\sigma)$. 
The probability of the optimal decoder (which applies a correction from the largest class) succeeding if $\sigma$ is observed is the probability the error came from that class
\begin{eqnarray}
p_{\text{opt}}(\sigma) = \frac{M_{1}(\sigma)}{M_{1}(\sigma)+M_{2}(\sigma)+M_{3}(\sigma)+M_{4}(\sigma)} \geq \frac{1}{4}. \nonumber
\end{eqnarray}
The random decoder selected error has a probability satisfying $p_{\text{rand}}(\sigma) \leq p_{\text{opt}}(\sigma)$ given explicitly by
\begin{eqnarray}
p_{\text{rand}}(\sigma) \!=\! \sum_{j=1}^4 \!\! \left(\frac{M_{j}(\sigma)}{\! M_{1}(\sigma) \!+\! M_{2}(\sigma) \!+\! M_{3}(\sigma) \! + \! M_{4}(\sigma)} \!\right)^{\!\! 2} \geq \frac{1}{4}, \nonumber
\end{eqnarray}
because there will be a probability proportional to $M_j$ that an error from class $j$ is applied, and then a probability proportional to $M_j$ that a correction will be applied (correcting the error).
These relations imply that
\begin{eqnarray}
\frac{1}{4} p_{\text{opt}}(\sigma)  \leq \frac{1}{4} \leq	p_{\text{rand}}(\sigma) \leq p_{\text{opt}}(\sigma) \leq 1.
\end{eqnarray}
This is true for all $\sigma$, which implies the following relations between the number of minimum weight failing configurations for the optimal and random minimum weight decoders:
\begin{eqnarray}
\frac{1}{4}N_{\text{fail opt}}^{\W}(d/2) \leq	N_{\text{fail rand}}^{\W}(d/2) \leq N_{\text{fail opt}}^{\W}(d/2).
\end{eqnarray}
%\mb{As $\gamma^{\W}$ is not sensitive to a multiplicative pre-factor in $N_{\text{fail}}^{\W}(d/2) \sim (\gamma^{\W})^{\sqrt{n}}$, it is the same for the random decoder as for the optimal decoder.}{}

%%%%%%%%%%%%%%%%%%%%%%%%%%%%%%%%%%%%%%%%%%%%

\section{Bounding $\gamma_0^{\W}$ explicitly}\label{app:pc}

Here we put explicit bounds on $\gamma_0^{\W}$ in $N_{\text{fail}}^{\W}(d/2) \sim (\gamma_0^{\W})^{\sqrt{n}}$.

{\it Upper bound:} First we consider the large $n$ limit of the upper bound for $N_{\text{fail}}^{\W}(d/2)$ in Eqn.~(\ref{eqn:pcW}) to upper bound $\gamma_0^{\W}$. 
First note that we can write Eqn.~(\ref{eqn:pcW}) as
\begin{eqnarray}
N_{\text{fail}}^{\W}(d/2) \leq N_{00} - N_{10}+N_{11},~~
\end{eqnarray} 
where,
\begin{eqnarray}
N_{ab} = d \sum_{T=0}^{d/4} \left( \! {C_{T-a}^{\frac{d}{2}-b} }\right)^{\!\!2} \sum_{w=0}^{w_{\text{max}}} 2^{T-w}  C_w^T C_{d/2 - T - w}^{d - 2T}.~~
\end{eqnarray} 
Using Stirling's approximation on all the Binomial terms, and replacing the sum by an integral,
\begin{eqnarray}
N_{ab} \rightarrow d \int_{0}^{d/4} \!\!\! \text{d}T \int_{0}^{r} \!\! \text{d}w \exp[f(T,w,d;a,b)],~~
\end{eqnarray}
where $f(T,w,d;a,b)$ is given by
\begin{eqnarray}
\!\!\!&&  (d - 2 T) \log\left[\frac{d - 2 T}{d}\right]  +(d - 2b)\log\left[\frac{d/2-b}{d}\right] + \nonumber  \\ 
&&- 2 (a - b + d/2 - T) \log\left[\frac{a - b + d/2 - T}{d}\right]  + \nonumber \\
&& -2 (T - a) \log\left[\frac{T-a}{d}\right] - (T - w)\log\left[\frac{T-w}{d}\right] + \nonumber \\
&& - (d/2 - T + w) \log\left[\frac{d/2 - T + w}{d}\right]  + T \log\left[\frac{T}{d}\right]+ \nonumber\\
&&- (d/2 - T - w) \log\left[\frac{d/2 - T - w}{d}\right] - w \log\left[\frac{w}{d}\right].~~~~~
\end{eqnarray}

First we consider the case $a=b=0$.
For all $d$, the function $f(T,w,d;0,0)$ is maximized for $w=d/18$ and $T = 2 d/9$.
We can expand $f$ about its maximum $f_0$ in powers of $\delta T = (T -\frac{2 d}{9})$ and $\delta w = (w -\frac{d}{18})$,
\begin{eqnarray}
&&f_0 \!+\! \frac{\partial^2 f}{\partial T^2}\frac{\delta T^2}{2} \!+\! \frac{\partial^2 f}{\partial T \partial w} \delta T \delta w \!+\! \frac{\partial^2 f}{\partial w^2}\frac{\delta w^2}{2} \!+\! \dots ,~~~\nonumber
\end{eqnarray}
where
\begin{eqnarray}
f_0&=& \frac{d}{2} \log\left[ \frac{27}{2} \right],\\
\frac{\partial^2 f}{\partial T^2}&=&-\frac{18}{d}=f_{TT},\\
\frac{\partial^2 f}{\partial T \partial w}&=&-\frac{9}{2d}=f_{Tw},\\
\frac{\partial^2 f}{\partial w^2}&=&-\frac{63}{2d}=f_{ww}.
\end{eqnarray}
For large $d$, since $1/\sqrt{f''}$ (for all second order differentials with respect to $T$ and $w$) grows slower with $d$ than the mean, $e^f$ becomes concentrated, and the integral becomes a two-dimensional Gaussian in the limit. 
We can also extend the limits of the integral without changing the value, and then calculate it exactly as a standard multi-dimensional Gaussian integral, 
\begin{eqnarray}
\!\!\!\!\!\! N_{00} \!\!\! &\rightarrow& \!\! d \! \cdot \!\!\! \int_{\!-\infty}^{\infty} \!\!\!\!\!\! \text{d}r \!\! \int_{-\infty}^{\infty} \!\!\!\!\!\!\!\! \text{d}w \exp \!\left[ \!f_{0} \! + \! \frac{f_{\!TT}\delta r^2 \! + \! 2 f_{\!Tw}\delta T \delta w \! + \!\! f_{\!ww}\delta w^2}{2} \right]\!\!, \!\!\!\!\!\!\!\! \nonumber \\
&=&\frac{2 \pi d }{\sqrt{f_{TT}f_{ww} -f_{Tw}^2} } \exp[f_0], \nonumber \\
&=& \frac{4 \pi d^2 }{27 \sqrt{3} } \exp\left( \frac{d}{2} \log\left[ \frac{27}{2} \right] \right).\label{eq:exponentialscalingwithd}
\end{eqnarray}

We can treat $N_{ab}$ for general $a$ and $b$ as we did for $N_{00}$, although we do not get quite as clean an analysis. 
As we are interested in large $d$ and $a=0,1$ and $b=0,1$ we can take the limit of small $a/d$ and $b/d$.
To lowest order in $a/d$ and $b/d$ we find that $f(T,w,d;a,b)$ is maximized for $w=\frac{d}{4} (-3 + \sqrt{9 - 16 T/d} + 4 T/d)$ and $T=2d/9 -56 b/(135d) + 14 a/(15d)$.
Although $f_{TT}$, $f_{Tw}$ and $f_{ww}$ can all be calculated to lowest order in $a/d$ and $b/d$, the only term which contributes to the exponential dependence of $N_{ab}$ on $d$ is $f_0$. 
To lowest order in $a/d$ and $b/d$ we find
\begin{eqnarray}
f_0&=& \frac{d}{2} \log\left[ \frac{27}{2} \right] - 2 b \log\left[ \frac{9}{5} \right] - 2 a \log\left[ \frac{5}{4} \right] +\mathcal{O}(1/d). \nonumber
\end{eqnarray}
Therefore we see that asymptotically, $N_{10}$ and $N_{11}$ scale exponentially with $d$ with precisely the same exponent as $N_{00}$ (but with different pre-factors). 
Hence $N_{00} - N_{10}+N_{11}$ scales with the same exponent too. i.e., as $d\rightarrow \infty$:
\begin{eqnarray}
\log \left[ N_{\text{fail}}^{\W}(d/2) \right]  &\leq & \frac{d}{2} \log\left[ \frac{27}{2} \right] + \mathcal{O}(\log[d]) .\label{eq:exponentialscalingwithd}
\end{eqnarray}
Recalling $d=\sqrt{n}$ and by comparison of Eqn.~(\ref{eq:exponentialscalingwithd}) with Eqn.~(\ref{eqn:pc2}), we can then read off $\gamma_0^{\W} \leq \sqrt{27/2} \approx 3.6742$.

{\it Lower bound:} Now we outline the proof of a lower bound for $\gamma_0^{\W}$. 
Consider a pair of weight-$d/2$ errors $E, E'$ such that $E'=L E$ for some logical operator $L$. 
Both $E$ and $E'$ will have the same sydrome, and therefore any decoder (which applies a correction based only on the syndrome) must fail on at least one of the errors.
Our strategy then is to lower bound the number of such pairs, where all elements of all pairs are unique.

Consider a horizontal path $L$ through coordinate $(0,0)$ with $T$ turns as described in Sec.~\ref{sec:pathcountingWen}.
Then consider the set of all errors such that each right turn has precisely one error.
There are $2^{T} C_{d/2 - T }^{d - 2T}$ such errors (corresponding to the $w=0$ term in Eqn.~\ref{Eqn:Turns}). 
Note that this set of errors is closed under multiplication by $L$: each error $E$ in the set is 'paired' with the other $E' = L E$.
Counting these for all horizontal paths we have the lower bound
\begin{eqnarray}
N_{\text{fail}}^{\W}(d/2) \geq M_{00} - M_{10}+M_{11},
\end{eqnarray} 
where
\begin{eqnarray}
M_{ab} = \sum_{T=0}^{d/4} \left( \! {C_{T-a}^{\frac{d}{2}-b} }\right)^{\!\!2} 2^{T} C_{d/2 - T }^{d - 2T}.~~
\end{eqnarray}
We do not include pre-factors of $d$ or vertical or diagonal errors as we do not seek a very tight bound.
In a very similar argument as used to establish the upper bound, one can take the large $n$ limit of this sum to give a Gaussian integral which can be evaluated. 
This leads to
\begin{eqnarray}
\log \left[ N_{\text{fail}}^{\W}(d/2) \right]  \geq  d \log[2 + \sqrt{2}] + \mathcal{O}(d),~~
\end{eqnarray}
which tells us that $\gamma_0^{\W} \geq 2 + \sqrt{2} \approx 3.4142$.

This lower bound is likely to be quite loose because it neglects all those error configurations which have any doubly occupied right turns.

\section{Monte Carlo data at large system sizes}
\label{App:MonteCarlo}

\begin{figure}
\includegraphics[width=\columnwidth]{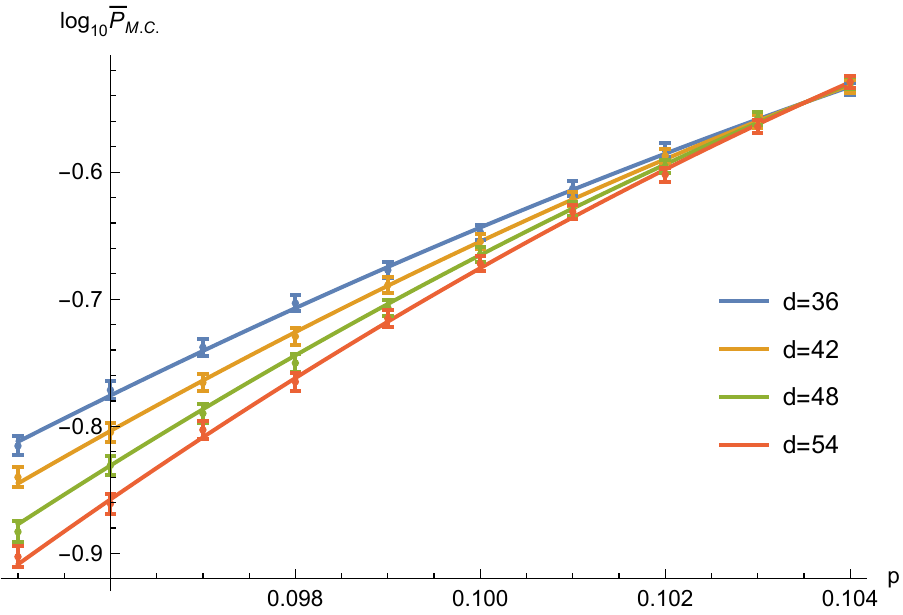}
\caption{Threshold fitted for the rotated-lattice model with system sizes $d =  36,\,42,\, 48,\,54$ using $N = 10^6$ samples for error rates $0.095 \le p \le 0.104$. The threshold value obtained is $p_{\text{th.}} = 0.1035 \pm 0.0002$. \label{Fig:WenThreshold}}
\end{figure}

In this Appendix we analyse the Monte Carlo data to find threshold error rates of the two models. We find the threshold error rates of these two models to be equivalent up to statistical fluctuations.
To determine the threshold we choose to collect data with error rate very close to the threshold error rate, $p_{\text{th.}} \sim 0.1 $, which has been determined previously~\cite{Dennis02}. We take data points for $ 0.095 \le p < 0.105$ as shown for the diamond lattice model in Fig.~\ref{Fig:WenThreshold}. We collect data for code distances $d = 20,\, 24, \, 28,\, 32$ with the square-lattice model.

We obtain a threshold value for each model by fitting the data to the function
\begin{equation}
\overline{P} = \exp(a + bx + cx^2), \label{Eqn:ThresholdAnsatz}
\end{equation}
with $x = (p-p_{\text{th}})^\mu$, where $a$, $b$, $c$, $\mu$ and $p_{\text{th}}$ are constants to be determined. The fitting parameters are given in Tab.~\ref{Tab:ThresholdFit}.
\begin{table}[b]
\begin{tabular}{|l|c|c|}
\hline
& $\K$ & $\W$ \\
\hline
\hline
$p_{\text{th}}$ & $0.1035\pm0.0002 $ &$ 0.1035 \pm 0.0002 $ \\ 
$ \mu $ & $ 0.68 \pm 0.03 $ & $ 0.63 \pm 0.03 $ \\ 
$a$ & $ -0.546 \pm 0.005 $ & $ -0.545 \pm 0.007 $ \\ 
$b$ & $ 2.9 \pm 0.2 $ & $ 2.7 \pm 0.3 $ \\ 
$c $& $ -7 \pm 1 $ & $ -7 \pm 2 $ \\
\hline
\end{tabular}
\caption{Parameters obtained by fitting the data collected with both the square and rotated model to Eqn.~(\ref{Eqn:ThresholdAnsatz}). The fitting is plotted to data for the rotated-lattice model in Fig.~\ref{Fig:WenThreshold}. \label{Tab:ThresholdFit}}
\end{table}

\begin{figure}
\includegraphics[width=\columnwidth]{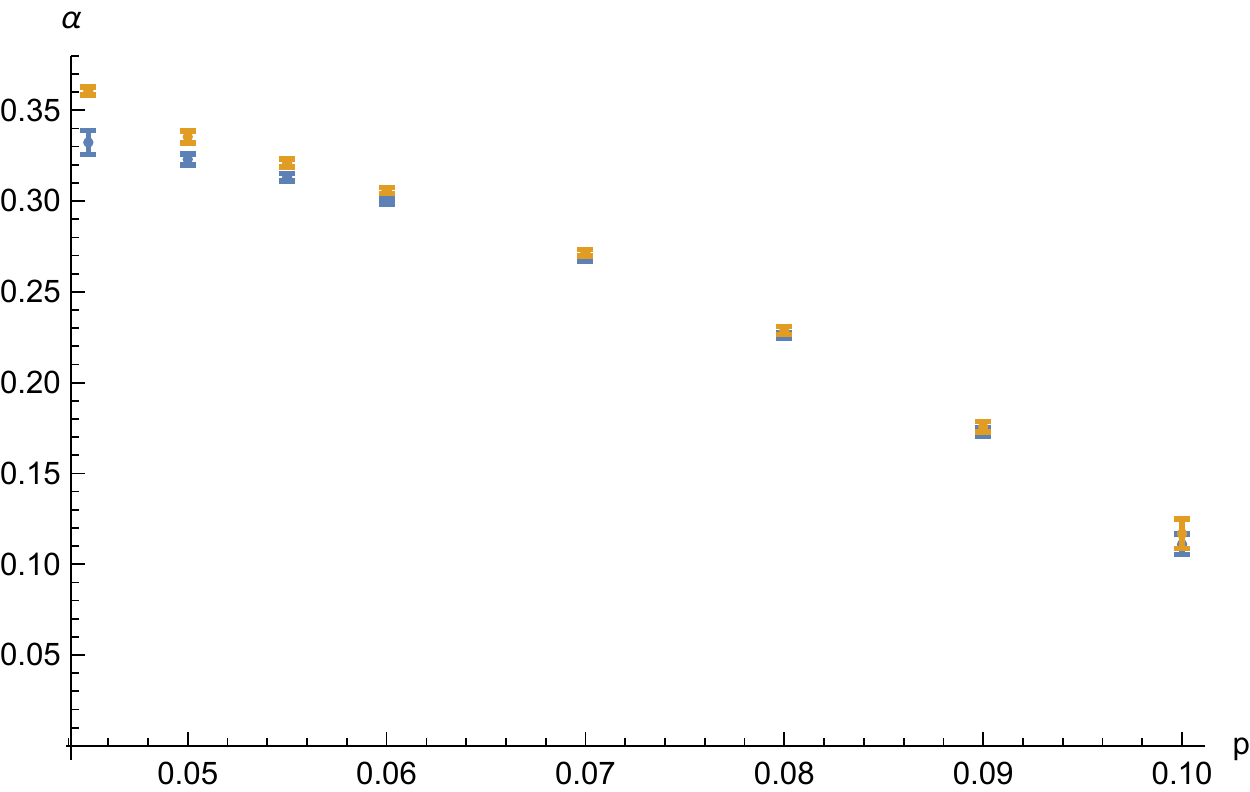}
\caption{\label{Fig:LargeAlpha} The values of $\alpha(p)$ obtained from the data in Fig.~\ref{Fig:LargeSizes}, where the data for square-lattice and rotated-lattice models are shown in blue and yellow respectively. The data is collected for relatively large sizes such that $40 \lesssim \sqrt{n} \lesssim 64$. The data shows that $\alpha(p)$ is greater for the rotated-lattice model for physical error rates in the interval $p_{\text{th.}}/ 2 \lesssim p \lesssim p_{\text{th.}}$, which suggests that if we extrapolate the logical failure rate to larger system sizes the rotated-lattice model will outperform the square-lattice model.}
\end{figure}

\subsection{Modelling logical failure rates}

\begin{figure}[b]
\includegraphics[width=\columnwidth]{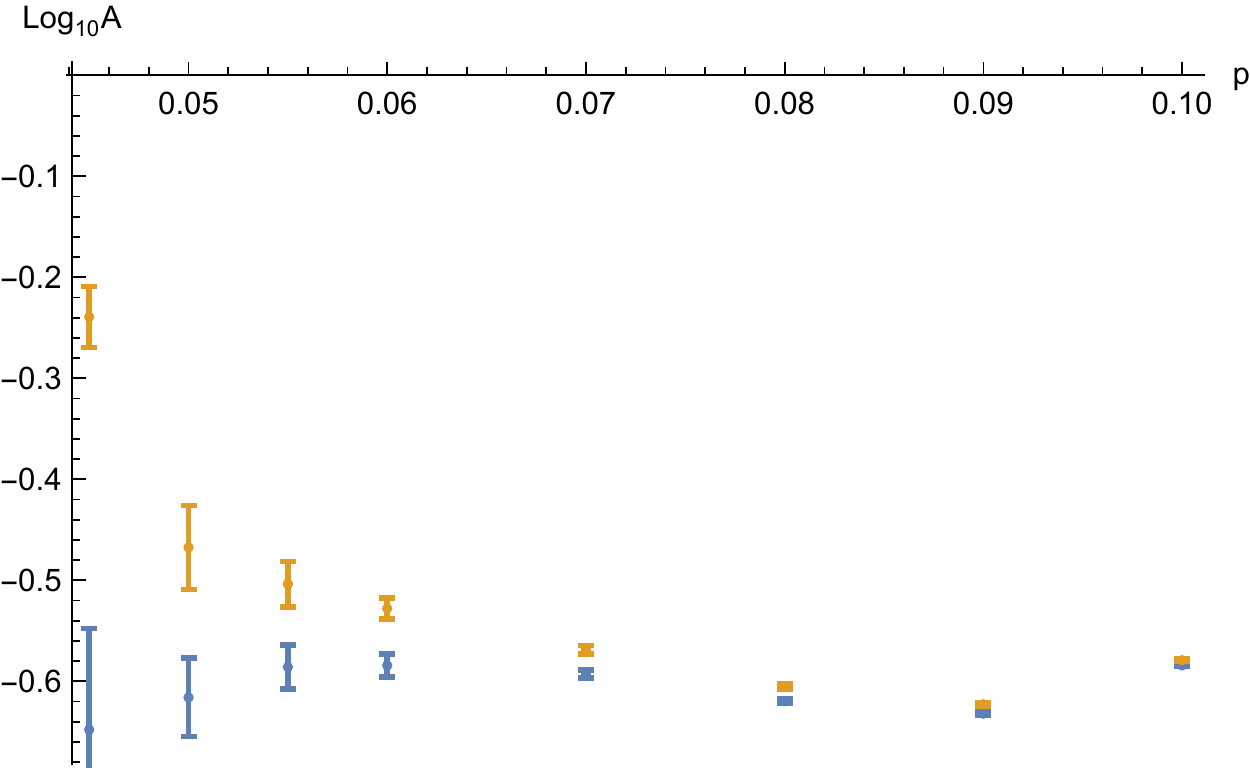}
\caption{\label{Fig:LargeLogA} Values of $\log_{10}A(p)$ for the square-lattice(rotated-lattice) surface code shown in blue(yellow) presented as a function of $p$. Data is collected for system sizes in the interval $40 \lesssim \sqrt{n} \lesssim 64$.}
\end{figure}

With the threshold error rates determined for each model, we fit to the Ansatz given in Eqn.~(\ref{Eqn:MCFittingAnsatz}) to determine $\alpha(p)$ and $\log_{10}A(p)$ for each value of $p$, as presented in Figs.~\ref{Fig:LargeAlpha} and~\ref{Fig:LargeLogA}, respectively. The data for the square-lattice(rotated-lattice) model is shown in blue(yellow). It is noteworthy that the data presented in these figures for large system sizes differs from the values of $\alpha(p)$ and $\log _{10}A(p)$ in Figs.~\ref{Fig:alpha} and~\ref{Fig:logA} where the values are determined using a different interval of system sizes. This shows that the assumption that $\alpha(p)$ and $\log_{10} A(p)$ are constant in $n$ is not valid in general. We find that the drift in these functions is slow in $n$, we thus assume that it is appropriate to fit a straight line to the data shown in Fig.~\ref{Fig:LargeSizes} provided we restrict to a relatively small interval of system sizes.

\section{Evaluating failure rates when logical errors are uncommon}
\label{App:Splitting}

Numerically sampling logical failure rates using standard Monte-Carlo methods are intractable below $ p \lesssim p_\text{th.} / 2$ since the likelihood of the minimum-weight matching decoder decays exponentially as $p$ decreases. It was the insight of Bravyi and Vargo in~\cite{Bravyi13} to use Bennett's acceptance ratio method~\cite{Bennett76} together with the splitting method, see eg. Ref.~\cite{Rubino09}, to numerically analyse the logical failure rate of the surface code as the physical error rate decreases towards the low error-rate regime. Here we briefly summarise the method of Bravyi and Vargo that was used to produce the low $p$ numerical results. 

The goal is to determine a logical failure rate 
\begin{equation}
\overline{P}(n, p_0)  = \sum_{E \in \mathcal{F}} \pi(E), \label{Eqn:LowFailureRate}
\end{equation} 
where $\mathcal{F}$ denotes the subset of errors that cause the decoder to fail for some system size $n$ and physical error rate $p_0$ that is intractable using Monte-Carlo techniques. To efficiently estimate Eqn.~(\ref{Eqn:LowFailureRate}) we first express the quantity of interest as a product of ratios, namely,
\begin{equation}
\overline{P}(n, p_0) = \overline{P}(n, p_\Lambda)  \prod_{j=0}^{\Lambda-1} \frac{\overline{P}(n, p_{j}) }{ \overline{P}(n, p_{j+1})}. 
\end{equation}
For brevity we write ratios $R_j = \overline{P}(n, p_{j}) /  \overline{P}(n, p_{j+1}) $ for a sequence of physical error rates $p_j$ for $0 \le j \le \Lambda$. Shortly, it will also become convenient to define
\begin{equation}
\pi_j(E) = (1-p_j)^{n - \text{wt}(\mathbf{x})} p_j^{\text{wt}(\mathbf{x})},
\end{equation}
similar to Eqn.~(\ref{Eqn:iidDistribution}) to specify the independent and identically distributed noise models for different error rates $p_j$ where $E = \prod_{j= 1}^n Z_j^{x_j}$ and $x_j = 0,\,1$ are elements of $\mathbf{x}$. Now, supposing $ \overline{P}(n, p_\Lambda)$ can be directly estimated because, say $p_\Lambda$ lies either in the path-counting regime, or is suitably high that we can use standard Monte Carlo methods, provided we have a method to estimate ratios $R_j$ for some sequence of physical error rates $p_j$ we can make estimates on $\overline{P}(n, p_0)$. We therefore turn to the acceptance ratio method due to Bennett to evaluate these ratios.

Bennett~\cite{Bennett76} showed that we can express the ratios we are trying to determine as follows
\begin{equation}
R_j = C \frac{ \left\langle g( C A_j) \right\rangle_j }{\left\langle g( C^{-1} A_j^{-1}) \right\rangle_{j+1} }, \label{Eqn:SplittingFraction}
\end{equation}
where the function $g(x) = 1 / (1+x)$ satisfies detailed balance, i.e., $g(x) = x^{-1} g(x^{-1})$, the term $C$ is a constant, we concisely write the ratio $A_j(E) = \pi_j(E) / \pi_{j+1}(E)$ and $ \langle f \rangle_j$ the expectation value of function $f(E)$ with respect to errors $E$ drawn from the probability distribution $\pi_j$ such that
\begin{equation}
\left\langle f \right\rangle_j = \frac{1}{Z} \sum_{E\in \mathcal{F}} f(E) \pi_j(E),
\end{equation}
and $Z = \sum_{E\in \mathcal{F}} \pi_j(E)$ normalises the probability distribution as we condition on errors that cause logical failures.

\begin{figure}
\includegraphics[width=\columnwidth]{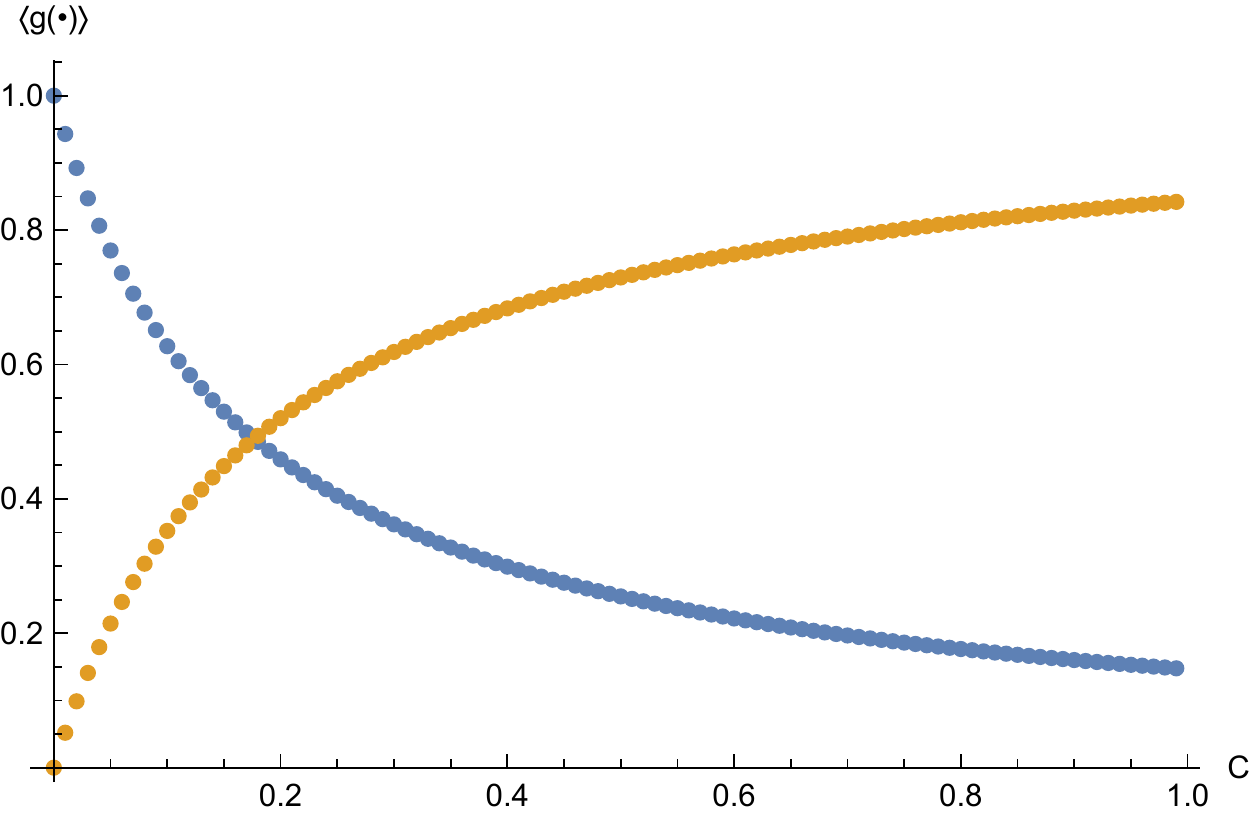}
\caption{\label{Fig:Rcalc} Generic plot showing $\langle g(C A_j ) \rangle_j$ in blue and $\langle g(C^{-1} A^{-1}_j ) \rangle_{j+1}$ in yellow plotted for 100 different values of $C$ between 0 and 1. The crossing point determines $C^*$.}
\end{figure}

We examine Eqn.~(\ref{Eqn:SplittingFraction}) to determine a method to find $R_j$. In particular we observe that for a particular value of $C = C^*$ such that $ \left\langle g( C^* A_j) \right\rangle_j  = \left\langle g( {C^*}^{-1} A_j^{-1}) \right\rangle_{j+1} $
whereby
\begin{equation} 
\frac{ \left\langle g( C^* A_j) \right\rangle_j  }{  \left\langle g( {C^*}^{-1} A_j^{-1}) \right\rangle_{j+1} } = 1,
\end{equation} 
we have that $R_j = C^*$. We therefore compare expectation values $\left\langle g( C A_j) \right\rangle_j$ and $ \left\langle g( C^{-1} A_j^{-1}) \right\rangle_{j+1}$ for different values of $C$ to determine $R_j$, see Fig.~\ref{Fig:Rcalc} for a generic demonstration of this method.

We evaluate the expectation values numerically, i.e.
\begin{equation}
\left\langle f \right\rangle_j \approx \sum_{\alpha = 1} ^N f(E_\alpha),
\end{equation}
for an arbitrary function $f$ of Pauli error $ E_\alpha $ that cause the decoder to fail where we choose errors according to the distribution $\pi_j(E)$. We draw error operators $E_\alpha$ from the correct distribution $\pi_j$ that cause the decoder to fail using the Metropolis-Hastings algorithm~\cite{Hastings70} where we use an error configuration $E_\alpha$ to determine $E_{\alpha+1}$.

\subsection{Implementation}

The procedure to find $E_{\alpha+1}$ in the algorithm we implement follows two steps. First, given an initial sample $E_\alpha$ we propose a new error operator $E_{\alpha}'$ by flipping a single qubit that is chosen uniformly over all the qubits of the system. Then, we decide to proceed with the proposed change with probability $\min (1, \pi_j(E_{\alpha}') / \pi_j(E_\alpha) )$. Otherwise, we continue with the original sample such that  $ E_{\alpha+1} = E_\alpha$. Given that we decide to proceed with $E_\alpha'$ we must then check that the trial error causes a logical failure. We therefore run the decoding algorithm to check the new error configuration causes the decoder to fail. We decide to accept the new error configuration such that $E_{\alpha+1} = E_\alpha ' $ only if the new trial error $E_{\alpha}'$ causes a logical failure. Otherwise we take $E_{\alpha+1} = E_\alpha$.

We propose $N = 10^9 $ new trials, and find that at least $ \sim 5 \cdot 10 ^5 $ different error configurations are accepted for each expectation value we calculate at distributions at the lowest values of $p$. For larger values of $p$ that we examine using the splitting method we may have as many as $\sim  10^8 $ error configurations that are sampled to determine expectation values.

\begin{figure}
\includegraphics[width=\columnwidth]{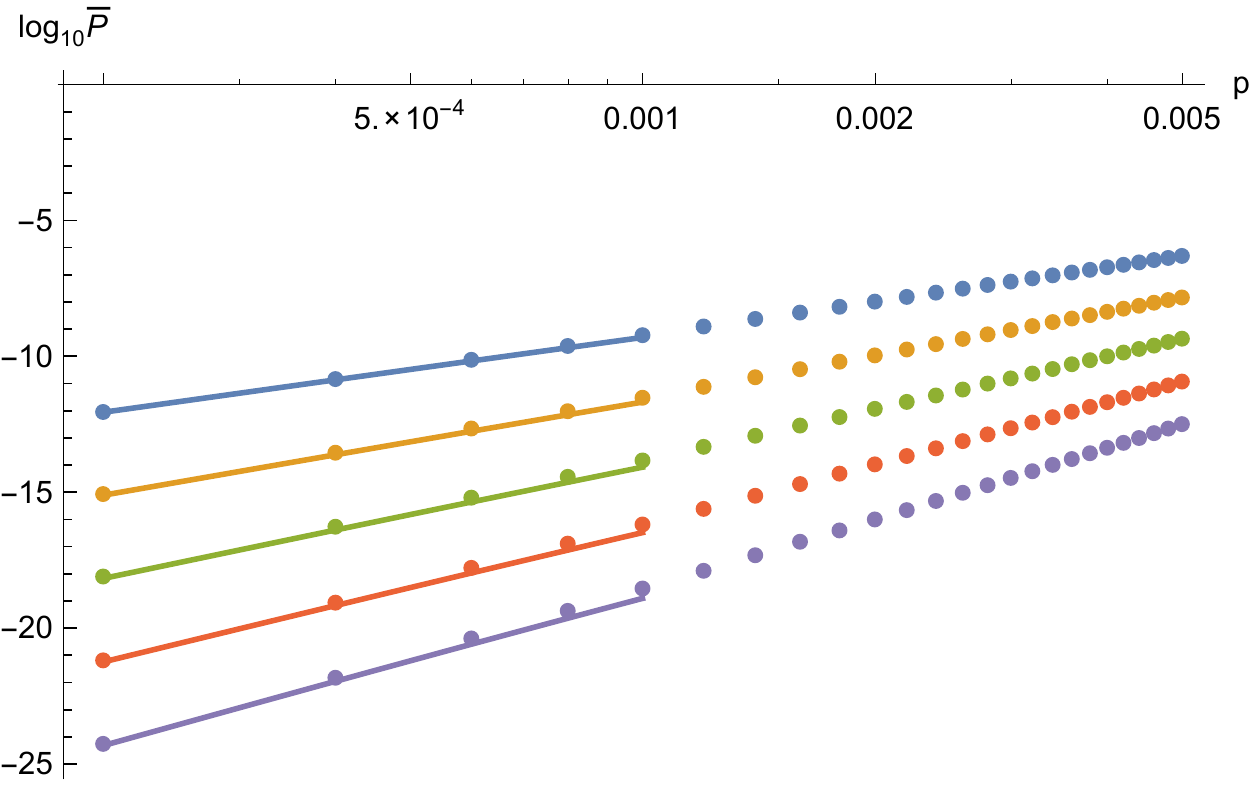}
\caption{Numerical data from the splitting method compared with the Kitaev path counting formula for system sizes $L = 8,\,10,\,12,\,14,\,16$. \label{Fig:KpathCountingData}}
\end{figure}

\begin{figure}
\includegraphics[width=\columnwidth]{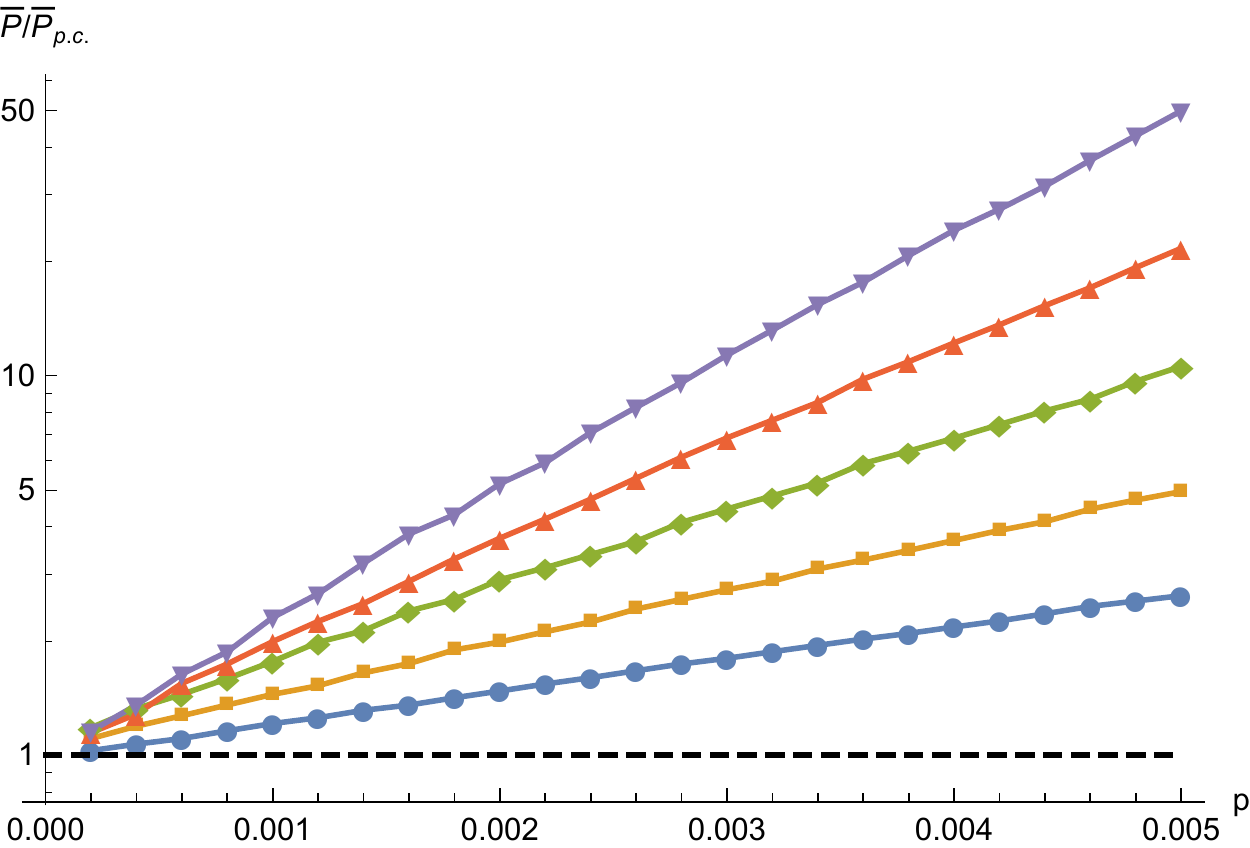}
\caption{Splitting method failure rates divided by the Kitaev path counting formula for system sizes $L = 8,\,10,\,12,\,14,\,16$. \label{Fig:KpathCountingRatios}}
\end{figure}

To illustrate the precision of our data points we compare the numerical data we collect using this method to the path-counting formula for the square lattice toric code. The data is shown in Fig.~\ref{Fig:KpathCountingData}. We observe our data points converging towards the first order approximation to the logical failure rate as $ p \rightarrow 0$. To show the convergence more clearly, in Fig.~\ref{Fig:KpathCountingRatios} we show the ratio of the logical failure rate and the path-counting approximation, $\overline{P}^\text{K} / \overline{P}_{\text{p.c.}}^\text{K}$. The convergence of the ratio towards $1$ as $p \rightarrow 0$ indicates the method produces accurate results.

\begin{figure}
\includegraphics[width=\columnwidth]{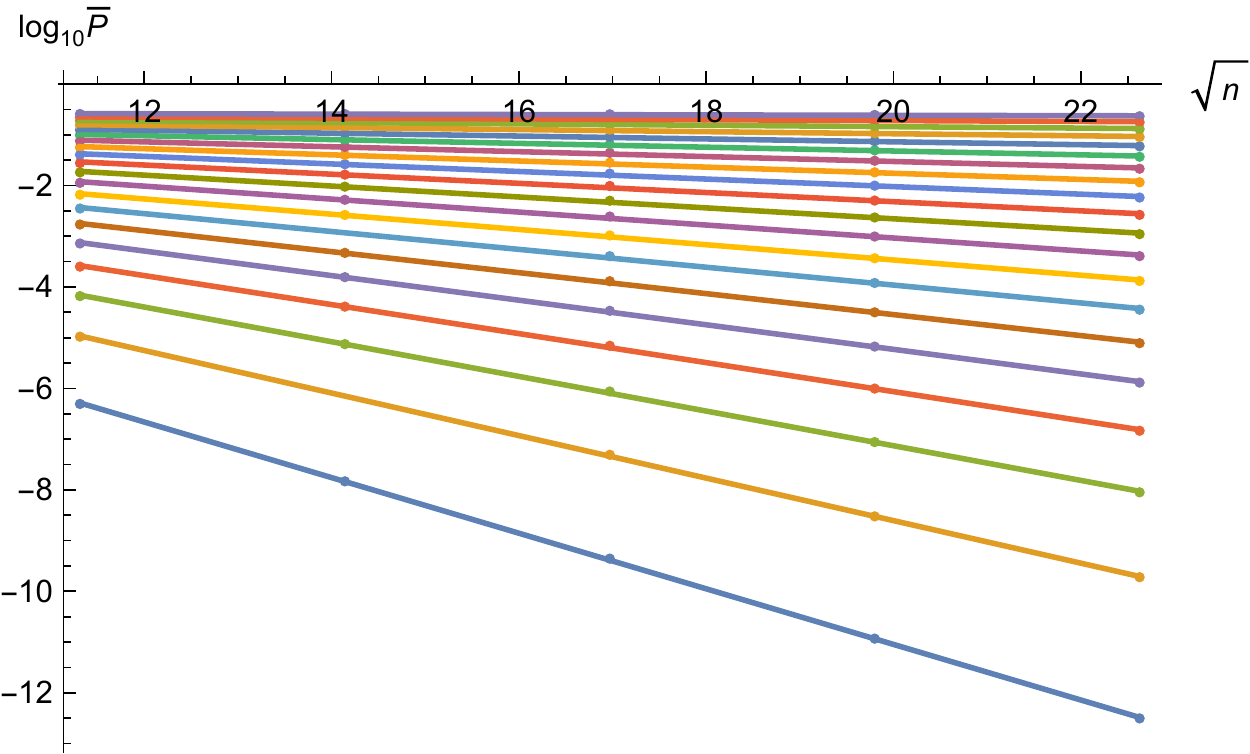}
\caption{Logical failure rates presented as a function of $\sqrt{n}$ for the square-lattice model. Gradients and intersections of the linear fittings are used to calculate the values of $\alpha(p)$ and $\log_{10}A(p)$ shown in Figs.~\ref{Fig:alpha} and~\ref{Fig:logA}, respectively. \label{Fig:KitSplitData}}
\end{figure}

\begin{figure}
\includegraphics[width=\columnwidth]{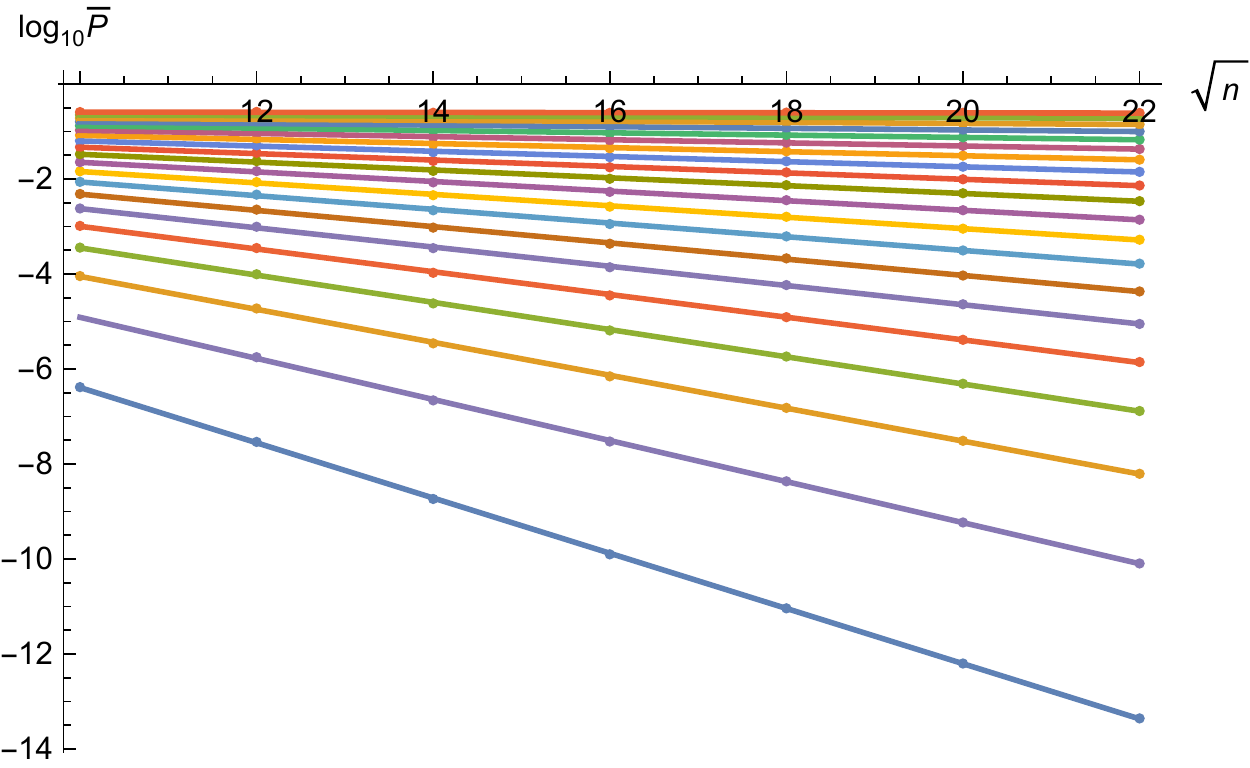}
\caption{Logical failure rates presented as a function of $\sqrt{n}$ for the rotated-lattice model. Gradients and intersections of the linear fittings are used to calculate the values of $\alpha(p)$ and $\log_{10}A(p)$ shown in Figs.~\ref{Fig:alpha} and~\ref{Fig:logA}, respectively. \label{Fig:WenSplitData}}
\end{figure}

We additionally calculate logical failure rates to determine $\alpha(p)$ and $\log_{10}A(p)$ in Figs.~\ref{Fig:alpha} and~\ref{Fig:logA} in the main text using the method due to Bravyi Vargo for physical error rates $p \le 5\%$. The data we use to find these graphs that is not displayed in Figs.~\ref{Fig:WenPathCounting} and~\ref{Fig:KpathCountingData} are shown in Figs.~\ref{Fig:KitSplitData} and~\ref{Fig:WenSplitData}. For error rates $p > 5\% $ we use Monte-Carlo methods.

%%%%%%%%%%%%%%%%%%%%%%%%%%%%%%%%%%%%%%%%%%%%

\section{Counting unconstrained non-contractible closed paths}
\label{app:unconstrainedpaths}

We estimate an asymptotic expression for the number of unconstrained paths as in Sec.~\ref{subsec:unconstrained}. We start with the expression

\begin{equation}
N_{\text{unc}}(l;x,y)= \!\!\!\sum_{n_\uparrow = y}^{(l + y - x)/2} \!\!\! \underbrace{\frac{l !}{n_\uparrow!  n_\downarrow(n_\uparrow)!  n_\rightarrow(n_\uparrow)!  n_\leftarrow(n_\uparrow)!}}_{\exp[B(n_\uparrow)]}, \!\!\label{eqn:NuncA}
\end{equation}
where $n_\downarrow=y+n_\uparrow$, $n_\leftarrow=\frac{1}{2}(l-2n_\uparrow-x-y)$ and $n_\rightarrow=\frac{1}{2}(l-2n_\uparrow+x-y)$.

In order to obtain a closed form approximation for Eqn. (\ref{eqn:NuncA}) valid in the large $n$ limit, we use Stirling's approximation to estimate, 

\begin{eqnarray}
B(n_\uparrow) &=& l \log l - n_{\uparrow} \log n_{\uparrow} - \log(n_{\uparrow} - y) \log[n_{\uparrow} - y]  \nonumber \\
&&-\frac{l - 2 n_{\uparrow} - x + y}{2} \log \left[\frac{l - 2 n_{\uparrow} - x + y}{2} \right]  \nonumber\\
&&- \frac{l - 2 n_{\uparrow} + x + y}{2} \log \left[\frac{l - 2 n_{\uparrow} + x + y}{2} \right] \nonumber\\
&&+O(\log{l}).
\end{eqnarray}

We now treat $l,x,y,n_\uparrow$ as continuous variables and expand $B(n_\uparrow)$ to second order about its maximum, resulting in a Gaussian sum, with a peak at $\mu$ such that $\frac{d}{d n_\uparrow} B(n_\uparrow) = 0$, and standard deviation $\sigma$ such that $\frac{d^2}{d n_\uparrow^2} B(n_\uparrow) = -1/(2 \sigma^2)$ evaluated at $n_\uparrow=\mu$. The mean and standard deviation are given by:

\begin{equation}
\mu  =  \frac{l^2 - x^2 + 2 l y + y^2}{4 l}, 
\end{equation}
and
\begin{equation}
\sigma =  \sqrt{\frac{l^4 + (x^2 - y^2)^2 - 2 l^2 (x^2 + y^2)}{8 l^3}}.
\end{equation}
Finally, we replace the sum over $n_\uparrow$ by an integral, and (since the Gaussian is concentrated far from the domain boundary) extend the limits to $\pm \infty$, to obtain
\begin{eqnarray}
\!\! N_{\text{unc}}(l;x,y) \!\! &\underset{\scriptscriptstyle l \rightarrow \infty}{\longrightarrow} & \!\! \int_{-\infty}^{\infty} d n_{\uparrow}  \exp \left[B(\mu)-\frac{(n_{\uparrow} - \mu)^2}{2 \sigma^2} \right]\!,~~\\
&=& \sqrt{2 \pi \sigma^2} \exp \left[B(\mu) \right].
\label{eq:gaussianintegral}
\end{eqnarray}

Substituting into the mean, we obtain a closed form approximation for $N_{\text{unc}}(l;x,y)$
\begin{eqnarray}
\frac{\log[N_{\text{unc}}]}{l} &\! = \! &  \log[4] + 2 \log[l] -  \frac{l \!+\! x \!+\! y}{2l} \log[l \!+\! x \!+\! y]+ \nonumber\\
&&\!- \frac{l \!+\! x \!-\! y}{2l} \log[l \!+\! x \!-\! y] 
\!-  \frac{l \!-\! x \!+\! y}{2l} \log[l \!-\! x \!+\! y]~~ \nonumber \\
&&- \frac{l \!-\! x \!-\! y}{2l} \log[l \!-\! x \!-\! y].
\end{eqnarray}

We plot $\log N_{\text{unc}}(l;x,y)$ for the two cases of interest in Fig.~\ref{fig:PlotUnconstrainedPaths}.
The curves for the two relevant orientations (i.e. $(x,y)=(\sqrt{n/2},0)$, and $(x,y)=(\sqrt{n}/2,\sqrt{n}/2)$) asymptotically approach one another for large $l/\sqrt{n/2}$.

We can see this clearly by expanding the closed expression for $\log N_{\text{unc}}(l;x,y)$ for large $l$ (and fixed $r/l<1$, with $r = \sqrt{x^2+y^2}$),
\begin{eqnarray}
\log N_{\text{unc}}(l;x,y) & = & l \log[4] - \frac{x^2+y^2}{l}  -\frac{(x^2+y^2)^2}{6l^3} + \nonumber\\
&&  -\frac{2x^2 y^2}{3l^3} + \mathcal{O}\left[r\left(\frac{r}{l}\right)^6\right].
\label{eq:unconstrainedpaths}
\end{eqnarray}
Note that $r=\sqrt{x^2+y^2}$ is identical for the two lattice rotations for fixed $n$, since $(x,y)=(\sqrt{n/2},0)$ for the standard lattice, and $(x,y)=(\sqrt{n}/2,\sqrt{n}/2)$ for the rotated lattice.

%Thanks to cluster computation we can collect data massively, and visualize the $n$ dependence of the codes for a given $p$: which is a more "physical" study, since the noise is fixed on an experimental system. Some results are given in Fig. \ref{pf_vs_n}: surprisingly, when $p$ is large (roughly between $1 \%$ and $10 \%$) both codes behave exactly the same. In the large $p$ regime, the sequences of points of Wen and Kitaev lattices overlap almost perfectly. However, a spitting occurs a low $p$ region, consistent with our previous study based on path counting hypothesis. 

%Even if we are not surprise by the low $p$ regime -- we know that due to the Wen distance, the failing rate must decay faster for the rotated lattice -- the equivalence of both lattices when $p$ is closer to threshold is unexpected.

%We would like to identify the region -- depending of $p$ and $n$ -- where the Wen lattice is clearly better the Kitaev one. A natural choice is plotting the ratio $P_f^W / P_f^K$ and the isolines of that ratio on a density plot. To achieve that we need to fit all the data collected by splitting method (since our values of $P_f$ are not evaluated for the same $p$ and $n$ parameters).

%\section{Further numerics}
%Introduce show and explain the binning results. Also show the two plots of $P_W/P_K$; numerics and analytics. 
%{\color{red} [[explain splitting method?discuss convergence?]]}

%%%%%%%%%%%%%%%%%%%%%%%%%%%%%%%%%%%%%%%%%%%%

\section{Estimating the finite size effects in the number of non-intersecting closed closed paths}
\label{sec:FiniteSizeNcl}

Here we describe how the asymptotic estimates of $N_{\text{con}}^\K(l,n)$ and $N_{\text{con}}^\W(l,n)$ were made in
 Fig.~\ref{fig:PlotUnconstrainedPaths} in Sec.~\ref{subsec:constrained}.

\begin{figure}[t]
	\includegraphics[width=\columnwidth]{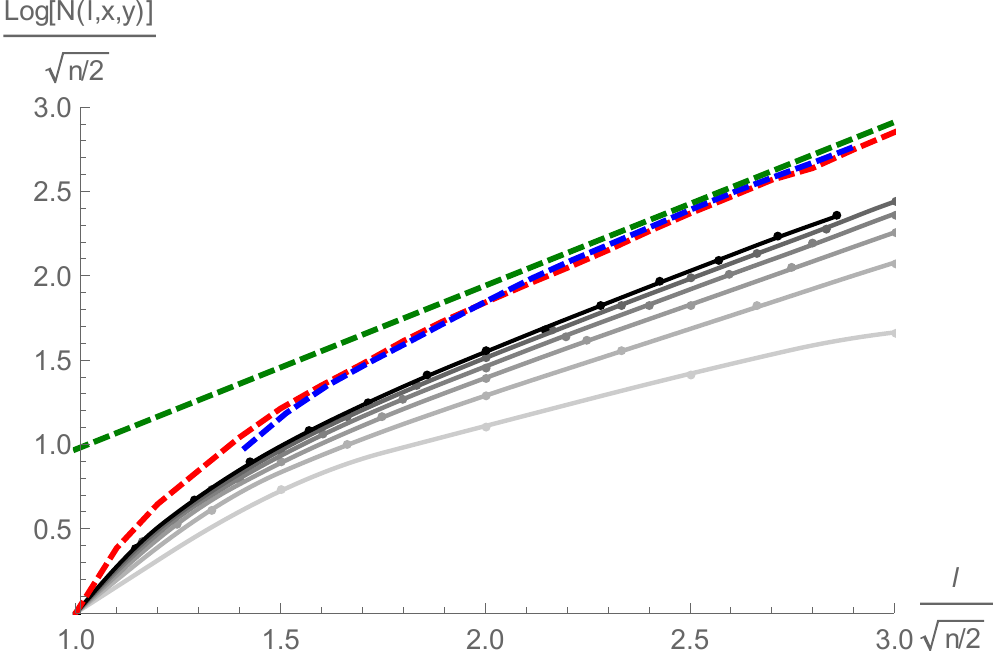}
	\caption{The number $N_{\text{con}}^{\K}(l,n)$ of constrained paths around the torus for $d=4,6,8, 10, 12, 14$ from light to dark. 
		For each data point, $N_{\text{con}}^\K(l,n)$ is found by sampling unconstrained paths and counting what fraction satisfy the constraints.
		The sample size is chosen to obtain 2\% accuracy.
		The red curve depicts the extrapolated $n \rightarrow \infty$ limit by fitting to ansatz Eq.~(\ref{eq:finitescaleansatz}).
		The blue curve is the limit for $N_{\text{con}}^{\W}(l,n)$ and is shown for reference.
		Both extrapolated curves appear to approach the bound $c^l$, with $c=2.638$ (green).}
	\label{fig:PlotConstrainedPathsKitaev}
\end{figure}

\begin{figure}[t]
	\includegraphics[width=\columnwidth]{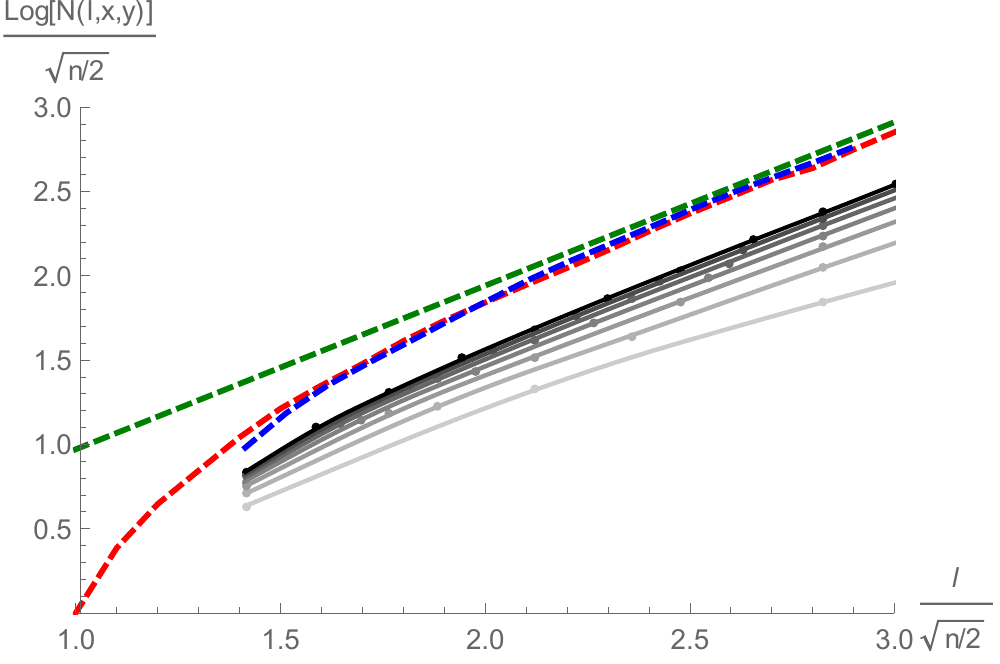}
	\caption{As for Fig.~\ref{fig:PlotConstrainedPathsKitaev} but for the rotated torus yielding $N_{\text{con}}^{\W}(l,n)$ for $d=4,6,8, 10, 12, 14$ from light to dark.
	}
	\label{fig:PlotConstrainedPathsWen}
\end{figure}

First, for each distance $d$, a curve $N_{\text{con}}(l,n)$ for $l=d, d+2, \dots, 3 \sqrt{n/2}$ was produced by randomly sampling among unconstrained closed closed paths and checking what fraction satisfied the non-intersection constraint.
The logarithm was taken and 
a smooth interpolation between the discrete data points was applied to produce a continuous function of $l$ for each $d$.
Then, from the set of such curves, for $\hat{l} = l/\sqrt{n/2}$ values between $1$ and $3$ in increments of $0.1$, we fit the following finite-size ansatz:
\begin{eqnarray}
\frac{\log \left[N_{\text{con}}\left(l ,n \right) \right]}{\sqrt{n/2}} = A(\hat{l}) -\frac{B(\hat{l}) }{\sqrt{n}} \log[C(\hat{l})  \cdot \sqrt{n}], \label{eq:finitescaleansatz}
\end{eqnarray} 
where $A(\hat{l})$, $B(\hat{l})$ and $C(\hat{l})$ are fitting parameters.
The value of $A(\hat{l})$ is taken to be the asymptotic limit of $\log [N_{\text{con}}]/\sqrt{n/2}$.
The data and the extrapolations are plotted for the standard lattice in Fig.~\ref{fig:PlotConstrainedPathsKitaev} and for the rotated lattice in Fig.~\ref{fig:PlotConstrainedPathsWen}.

This ansatz fits the finite size data well and is additionally justified by our analytic understanding of the following non-trivial example which we can analyze exactly. 
For the rotated lattice at $l=d =\sqrt{n}$, we know from Sec.~\ref{subsec:rotatedpathcounting} that,
\begin{eqnarray}
N_{\text{con}}^\W\left(\sqrt{n} ,n \right) &=& C^{d}_{d/2} =  \frac{d!}{(\frac{d}{2})!(\frac{d}{2})!},\nonumber\\
\frac{\log\left[ N_{\text{con}}^\W\left(\sqrt{n} ,n \right) \right]}{\sqrt{n/2}} &=& \frac{\sqrt{2}}{d}\log \left[ \frac{d!}{(\frac{d}{2})!(\frac{d}{2})!} \right],\nonumber\\
& \longrightarrow & \frac{\sqrt{2}}{d}\log \left[ \frac{\sqrt{2 \pi d}}{2 \pi \frac{d}{2}} \left(\frac{d}{e}\right)^{d} \left(\frac{d}{e}\right)^{-d} \right],\nonumber\\
& = & \sqrt{2} \left[ \log[2] - \frac{1}{2\sqrt{n}} \log(8 \pi \sqrt{n}) \right].\nonumber
\end{eqnarray} 
This is precisely the form of the ansatz, and the asymptotic value $\sqrt{2} \log[2] = 0.98026 \dots$ is close to that ($0.97$) found from fitting to the ansatz for just seven distances $d=4,6,8,10,12,14$.

\bibstyle{plain}

\end{document}